\documentclass[11pt]{article}
\usepackage[a4paper, margin=1in]{geometry}
\usepackage[sorting = anyt, maxbibnames=9,maxcitenames=2, backend = bibtex, style=authoryear, doi=false, url=false]{biblatex}

\renewbibmacro{in:}{}
\usepackage[titletoc,page]{appendix}

\usepackage{soul}
\usepackage{graphicx}
\usepackage{setspace}
\usepackage{gensymb}
\usepackage{enumerate}
\usepackage{kotex}
\usepackage{geometry}
\usepackage{abstract,lipsum}
\usepackage{color}
\usepackage{amssymb}
\usepackage{amsmath, nccmath}
\usepackage{amsthm}
\usepackage{amsfonts}
\usepackage{upgreek}
\usepackage{mathtools}
\usepackage{mathrsfs}
\usepackage{bm}
\usepackage{caption}
\usepackage{subcaption}
\usepackage{indentfirst}

\usepackage[symbol]{footmisc}

\DeclareMathAlphabet\mathrsfso{U}{rsfso}{m}{n}
\DeclareMathAlphabet\mathbfcal{OMS}{cmsy}{b}{n}
\usepackage[colorlinks=true,linkcolor=blue,citecolor=blue,urlcolor=blue]{hyperref}
\usepackage{array}

\usepackage{enumitem}
\SetLabelAlign{Center}{\hfil#1\hfil}
\SetLabelAlign{CenterWithParen}{\hfil(\makebox[1.0em]{#1})\hfil}

\begin{document}
\begin{center}
       \fontsize{19pt}{19pt}\selectfont Deformation and dislocation evolution in body-centered-cubic single- and polycrystal tantalum
       \vspace*{0.3in}

       \fontsize{10pt}{10pt}\selectfont Seunghyeon Lee, Hansohl Cho$^{\dagger}$\\
       \vspace*{0.05in}
       \fontsize{9pt}{9pt}\selectfont Department of Aerospace Engineering, Korea Advanced Institute of Science and Technology, Daejeon, 34141, Republic of Korea \\ 
       \vspace*{0.3in}
       \fontsize{10pt}{10pt}\selectfont Curt A. Bronkhorst$^{\ddagger}$\\
       \vspace*{0.05in}
       \fontsize{9pt}{9pt}\selectfont Department of Engineering Physics, University of Wisconsin, Madison, WI 53706, USA \\
       \vspace*{0.3in}
       \fontsize{10pt}{10pt}\selectfont Reeju Pokharel, Donald W. Brown, Bj$\o$rn Clausen, Sven C. Vogel, Veronica Anghel, George T. Gray \uppercase\expandafter{\romannumeral3}\\
       \vspace*{0.05in}
       \fontsize{9pt}{9pt}\selectfont Materials Science \& Technology Division, Los Alamos National Laboratory, Los Alamos, NM 87545, USA \\
       \vspace*{0.3in} 
       \fontsize{10pt}{10pt}\selectfont Jason R. Mayeur\\
       \vspace*{0.05in}
       \fontsize{9pt}{9pt}\selectfont Mechanical and Aerospace Engineering Department, University of Alabama in Huntsville, Huntsville, AL 35899, USA \\
\end{center}
\vspace*{0.1in}
\fontsize{8pt}{8pt}\selectfont E-mails: $^{\dagger}$ hansohl@kaist.ac.kr (Hansohl Cho), $^{\ddagger}$ cbronkhorst@wisc.edu (Curt A. Bronkhorst)
\renewenvironment{abstract}
{\small
\noindent \rule{\linewidth}{.5pt}\par{\noindent \bfseries \abstractname.}}
{\medskip\noindent \rule{\linewidth}{.5pt}
}

\vspace*{0.1in}
\onehalfspacing
\begin{abstract}
\fontsize{11pt}{11pt}\selectfont
A physically-informed continuum crystal plasticity model is presented to elucidate the deformation mechanisms and dislocation evolution in body-centered-cubic (bcc) tantalum widely used as a key structural material for mechanical and thermal extremes. We show our unified structural modeling framework informed by mesoscopic dislocation dynamics simulations is capable of capturing salient features of the large inelastic behavior of tantalum at quasi-static (10$^{-3}$ s$^{-1}$) to extreme strain rates (5000 s$^{-1}$) and at room temperature and higher (873K) at both single- and polycrystal levels. We also present predictive capabilities of our model for microstructural evolution in the material. To this end, we investigate the effects of dislocation interactions on slip activities, instability and strain-hardening behavior at the single crystal level. Furthermore, \textit{ex situ} measurements on crystallographic texture evolution and dislocation density growth are carried out for the polycrystal tantalum specimens at increasing strains. Numerical simulation results also support that our modeling framework is capable of capturing the main features of the polycrystal behavior over a wide range of strains, strain rates and temperatures. The theoretical, experimental and numerical results at both single- and polycrystal levels provide critical insight into the underlying physical pictures for micro- and macroscopic responses and their relations in this important class of refractory bcc materials undergoing severe inelastic deformations.\\
\end{abstract}

\doublespacing
\section{Introduction}
Body-centered-cubic (bcc) crystalline tantalum is a refractory transition metal in the Group \uppercase\expandafter{\romannumeral5}. It has been widely used for key structural components often exposed to harsh physico-chemical environments due to its superb strength, ductility and corrosion and radiation resistance over a wide range of strains, strain rates and temperatures. The mechanical behavior of tantalum and its alloys has been a focal point of research to facilitate their applications in diverse mechanical, thermal and chemical extremes. Recently, these materials are also finding new avenues towards high performance metallic composites and laminates for biological, defense and energy applications (\cite{Matsuno2001, Pappu1996, Mayeur2013}).

 Inelasticity in bcc materials has been a long-standing interest. Inelastic deformation mechanisms in bcc materials are fundamentally different from those in face-centered-cubic (fcc) or hexagonal materials. Inelastic slip in bcc materials are governed mainly by the motion of screw dislocations in a close-packed direction of $\langle$111$\rangle$  on several different planes of $\{$110$\}$, $\{$112$\}$ or $\{$123$\}$. As pointed out in \cite{Lim2020}, the planes where slip occurs in bcc crystals have been deemed elusive. Furthermore, inelastic slip of screw dislocations is a thermally-activated process over a wide range of strain rates and temperatures as well postulated in \cite{Stainier2002} and \cite{Nguyen2021}. The high Peierls stress (or lattice friction) associated with screw dislocations and the thermally-activated formation of kink pairs give rise to strongly rate- and temperature-dependent inelastic features in bcc crystals and their alloys. Over the past several decades, the rate- and temperature-dependent inelasticity has been investigated for single crystalline tantalum (\cite{Byron1968, Stainier2002, Lim2020, Nguyen2021}) and polycrystalline tantalum (\cite{Hoge1977, Kothari1998, Nemat-Nasser1998}).

In conjunction with the experimental and theoretical studies on the rate- and temperature-dependent inelastic features due to the complex motion of screw dislocations, the breakdown of the classical Schmid law also known as non-Schmid effects has been widely reported for bcc materials since the classical work by \cite{Taylor1926} and \cite{Taylor1928}. A non-planar core structure of the dominant screw dislocations gives rise to such abnormal plasticity features as tension-compression asymmetry and orientation-dependent critical shear stress for the onset of inelastic slip as experimentally evidenced in \cite{Byron1968} and \cite{Sherwood1967}. Moreover, the underlying physics of the non-planar core structure in $\langle$111$\rangle$ screw dislocations has been studied via atomistic simulations (\cite{Duesbery1973, Groger2008a, Groger2008b, Yang2001}) and ab-initio calculations (\cite{Ismail-Beigi2000, Dezerald2016}). Manifestation of non-Schmid effects has been well evidenced in experiments and atomistic simulations for tantalum materials, especially at low temperature (\cite{Dezerald2015, Sherwood1967}).

Meanwhile, classical continuum models of single crystalline materials are traced back to the seminal work by \cite{taylor1938plastic}, \cite{Hill1972}, \cite{Asaro1983} and \cite{Asaro1985}. More recently, the finite deformation single crystal plasticity theory was further developed within a thermodynamically consistent framework involving both macroscopic- and microscopic force balances (\cite{Anand2004, Gurtin2000}). Moreover, robust implicit numerical procedures for updating the constitutive models within finite deformation framework have been well established for use in nonlinear finite elements for boundary value problems of single- and polycrystal materials (\cite{Kali, Cuitino1993, Miehe1999}). Based upon these classical papers, numerous single crystal plasticity models have been proposed and have found success in elucidating the key features in the mechanical behaviors of bcc single crystals including tantalum (\cite{Nguyen2021}), $\alpha$-iron (\cite{Narayanan2014}), niobium (\cite{Mayeur2013}), tungsten (\cite{Cereceda2016}) and their alloys (\cite{Chen1996}) under diverse loading scenarios over a wide range of crystallographic orientations. Moreover, as recently reviewed in \cite{Cho2018}, continuum single crystal plasticity models have been extended to capture the non-Schmid behavior of tantalum and other bcc materials, especially at low temperatures, informed from atomistic simulations on the non-planar core structure of an isolated screw dislocation under diverse loading conditions. 

Single crystal plasticity theories have also enabled accurate modeling of the mechanical behavior of polycrystalline materials via nonlinear finite elements. The early models of polycrystalline behaviors involving the evolution of anisotropy due to crystallographic texturing date back to the work by \cite{Asaro1983} and \cite{Asaro1985} based on the classical work by \cite{taylor1938plastic}, in which all grains were assumed to have an equal volume and the deformation gradient within each grain was assumed to be uniform throughout the aggregate of the grains. Although these Taylor-type polycrystal models do not satisfy “compatibility” throughout the grain aggregate network, they have found success in part in modeling the nonlinear mechanical behavior and the evolution of crystallographic textures during deformation in many fcc (\cite{Asaro1985, Kali, Miehe1999}) and bcc materials (\cite{Kothari1998, Nemat-Nasser1998}). Polycrystal models have been recently extended to satisfy both equilibrium and geometric compatibility, for which each of the finite elements represents one single crystalline grain (\cite{Kali, Bronkhorst1992, Anand2004}). Furthermore, recent progress in electronic backscatter diffraction analysis of actual polycrystalline microstructures has enabled better modeling of the polycrystalline behavior with more realistic microstructures and networks of grains and grain boundaries especially for polycrystal bcc materials and laminates (\cite{Knezevic2014}). However, in these papers, the details regarding constituent single crystal behavior have not been presented independently of the polycrystal behavior; i.e., the previous modeling efforts have been focused mainly on the effective responses of the polycrystals. Furthermore, as the extreme thermomechanical responses associated with shear band localization, ductile damage and spallation upon harsh loading events have recently received great attention (\cite{Bronkhorst2016, , Bronkhorst2021, Kraus2021}), there has been an increasing need for unified structural modeling frameworks for both tantalum single- and polycrystals at a wide range of strains, strain rates and temperatures.

This work aims at elucidating deformation mechanisms and dislocation structure evolution in bcc single- and polycrystal tantalum using a suite of theoretical modeling, numerical simulation and experimentation. We present a physically-informed finite deformation single crystal viscoplasticity model in which the underlying physics of dislocation evolution and interaction throughout the slip systems is taken into account. Then, we show the predictive capabilities of the single crystal model at strain rates ranging from 0.001 s$^{-1}$ to 10$^3$ s$^{-1}$ and at temperatures ranging from room temperature to 873K for various crystallographic orientations. The model is further validated for the inelastic features in polycrystal tantalum. To this end, we conducted an extensive set of new experiments for the polycrystal specimens including mechanical tests and \textit{ex situ} neutron diffraction measurements on dislocation density growth and texture evolution, and compared the experimental data with the corresponding numerical simulation results. Using the unified modeling framework for both single- and polycrystal tantalum, we further develop critical insights into the inelastic deformation mechanisms at both microscopic- and macroscopic levels in this important class of refractory bcc materials.

The main body of this paper is as follows. The single crystal plasticity model is presented together with results on the single crystal behavior of tantalum from low- to high strain rates and at room temperature and higher in Section~\ref{section:single_cry}. Then, the single crystal model is further validated for polycrystal behaviors in both experiments and numerical simulations in Section \ref{section:polycrystal}, where we directly compare experimental data for evolution of texture and dislocation density and macroscopic mechanical responses against numerical results. Finally, we briefly summarize and discuss our main conclusions on the deformation mechanisms and the dislocation evolution in single- and polycrystal tantalum at various loading conditions in Section \ref{section:discussion}. Furthermore, some numerical details used in our analysis are provided in the Appendix.

A complete list for mathematical symbols used throughout this work is given in Table \ref{table:slip_symbol}.

\begin{table}[h!]
\centering
\renewcommand{\arraystretch}{1.2}
\caption{List of symbols.}
\begin{tabular}{ m{2.5cm}  m{12cm}}
\hline
Symbol                & Definition or meaning \\
\hline
$\mathbf{F}, \, \mathbf{F}^{\textrm{e}}, \, \mathbf{F}^{\textrm{p}}$ & Total, elastic, and plastic deformation gradients\\
$\mathbf{R}^{\textrm{e}}, \, \mathbf{U}^{\textrm{e}}$ & Elastic rotation, right stretch\\
$\mathbf{C}^{\textrm{e}}, \, \mathbf{E}^{\textrm{e}}$ & Elastic right Cauchy-Green tensor, elastic strain tensor   \\
$\mathbf{L}, \, \mathbf{L}^{\textrm{e}}, \, \mathbf{L}^{\textrm{p}}$ & Velocity gradient, elastic and plastic distortion rates\\
$\dot{\gamma_\textrm{p}}^{\alpha}, \, \dot{\gamma_0}$   & Plastic shear strain rate in the slip system $\alpha$, reference slip rate\\
$\mathbf{m}^{\alpha}_0, \, \mathbf{n}^{\alpha}_0$ & Slip direction and slip plane normal for slip system $\alpha$ \\
$\mathbf{\mathbb{S}}^{\alpha}_0, \, \mathbf{\mathbb{S}}^{\alpha}$     & Schmid tensors in intermediate space and deformed configuration\\
$\Phi $                & Elastic free energy \\
$\bm{\mathcal{C}} $                & Fourth order elastic stiffness\\
$\mathcal{C}_{11}, \, \mathcal{C}_{12}, \,  \mathcal{C}_{44}$    & Elastic constants at current temperature \\
$\mathcal{C}_{11,0}, \, \mathcal{C}_{12,0}, \,  \mathcal{C}_{44,0}$    & Elastic constants at 0K  \\
$m_{11}, \, m_{12}, \, m_{44}$    & Slopes of the temperature-dependent elastic constants \\
$\mathbf{A}$                & Thermal expansion tensor             \\
$\theta, \, \theta_0$ & Current and reference absolute temperatures\\
$\mu, \, \mu_0$ & Effective shear moduli at current temperature and 0K\\
$ \mathbf{T}^e, \, \mathbf{P}, \, \mathbf{T}$ & Elastic 2nd Piola stress, Piola stress, Cauchy stress\\
$\tau^{\alpha}, \, \tau^{\alpha}_{eff} $ & Resolved shear stress, effective shear stress in slip system $\alpha$ \\
$\Delta G$            & Activation energy \\
$k_B$                 & Boltzmann's constant \\
$p, \, q$             & Parameters for the shape of stress-dependent kink-pair formation energy\\
$s_0$             & Far-field slip resistance\\
$s^{\alpha}$             & Slip resistance in slip system $\alpha$\\
$\widetilde{s_l}, \,s_l$ & Temperature-dependent lattice resistance, lattice resistance at 0K \\
$b$ & Magnitude of Burgers vector\\
$a^{\alpha \beta}$    & Dislocation interaction matrix\\
$\rho^{\alpha}$    & Dislocation density in slip system $\alpha$ \\
$\mathcal{L}^{\alpha}$    & Mean free path of dislocation for slip system $\alpha$\\
$k_1, \, k_2$ & Mean free path coefficients\\
$y_c^{\alpha}, \, y_{c0}$    & Annihilation capture radius for slip system $\alpha$, reference capture radius\\
$A_{rec}$    & Capture radius energy\\
$\rho$    & Material mass density\\
$c$    & Specific heat\\
$\eta$    & Taylor-Quinney factor\\
\hline
\end{tabular}
\label{table:slip_symbol}
\end{table}
\clearpage

\section{Single crystal behavior}
\label{section:single_cry}

\subsection{Single crystal plasticity model}
\label{section:single_plasticity}

\subsubsection{Kinematics}
\label{section:kinematics}
The deformation gradient is defined by,
\begin{equation}
\mathbf{F}=\textrm{Grad} \, \mathbf{y},\\ 
\end{equation}
where $\mathbf{y} = \boldsymbol{\varphi}(\mathbf{X}, t)$ is the spatial vector mapped via the motion, $\boldsymbol{\varphi}$ and $\mathbf{X}$ is the material vector in the reference configuration. Here, ``Grad'' denotes a gradient in the reference configuration. The deformation gradient multiplicatively decomposes into its elastic ($\mathbf{F}^{\textrm{e}}$) and plastic ($\mathbf{F}^{\textrm{p}}$) parts, 
\begin{equation}
\mathbf{F}=\mathbf{F}^{\textrm{e}}\mathbf{F}^{\textrm{p}}. \\ 
\end{equation}

The spatial velocity gradient represents the rate of deformation in the deformed configuration by,
\begin{equation}
\begin{aligned}
& \mathbf{L} = \textrm{grad} \, \mathbf{v} =\dot{\mathbf{F}}\mathbf{F}^{-1},
\end{aligned}
\end{equation}
where $\mathbf{v}$ is the spatial velocity field and ``grad'' is a gradient in the deformed configuration. The velocity gradient additively decomposes into elastic ($\mathbf{L}^{\textrm{e}}$) and plastic ($\mathbf{L}^{\textrm{p}}$) distortion rate tensors, 
\begin{equation}
\mathbf{L} =\mathbf{L}^{\textrm{e}} + \mathbf{F}^{\textrm{e}}\mathbf{L}^{\textrm{p}}\mathbf{F}^{\textrm{e} -1},\\
\label{eqn:vg}
\end{equation}
where $\mathbf{L}^{\textrm{e}}=\dot{\mathbf{F}}^{\textrm{e}}\mathbf{F}^{\textrm{e} -1}$ and $\mathbf{L}^{\textrm{p}}=\dot{\mathbf{F}}^{\textrm{p}}\mathbf{F}^{\textrm{p} -1}$. Rearranging Equation (\ref{eqn:vg}), we have,
\begin{equation}
\dot{\mathbf{F}}^{\textrm{p}}=\mathbf{L}^{\textrm{p}}\mathbf{F}^{\textrm{p}}\;\textrm{with} \; \mathbf{F}^{\textrm{p}}(\mathbf{X},0) = \mathbf{1}.\\
\end{equation}

The dislocation motion is assumed to take place throughout prescribed slip systems $\alpha\, = 1 \sim N$ in the lattice space. Here, we define the Schmid tensor $\mathbf{\mathbb{S}}^{\alpha}_0=\mathbf{m}^{\alpha}_0 \otimes \mathbf{n}^{\alpha}_0$, where $\mathbf{m}^{\alpha}_0$ is the slip direction and $\mathbf{n}^{\alpha}_0$ is the slip plane normal defined in the intermediate (or lattice) space elastically relaxed from the deformed spatial configuration. Since plastic flow takes place throughout the prescribed slip systems, the plastic distortion rate tensor in the lattice space is expressed by,
\begin{equation}
\mathbf{L}^{\textrm{p}} = \sum_{\alpha=1}^{N} \dot{\gamma_\textrm{p}}^{\alpha} \mathbf{\mathbb{S}}^{\alpha}_0. \\
\end{equation}
Here, $\dot{\gamma_\textrm{p}}^{\alpha}$ is the plastic shear strain rate in each of the slip systems.
Equivalently, the rate of plastic distortion is expressed in the deformed configuration by,
\begin{equation}
\overline{\mathbf{L}^{\textrm{p}}}=\mathbf{F}^{\textrm{e}}\mathbf{L}^{\textrm{p}}\mathbf{F}^{\textrm{e} -1} = \sum_{\alpha=1}^{N} \dot{\gamma_\textrm{p}}^{\alpha} \mathbf{\mathbb{S}}^{\alpha},\\
\end{equation}
where $\mathbf{\mathbb{S}}^{\alpha} = (\mathbf{F}^{\textrm{e}}\mathbf{m}^{\alpha}_0) \otimes (\mathbf{F}^{\textrm{e}\textrm{-T}}\mathbf{n}^{\alpha}_0)$ is the Schmid tensor in the deformed configuration.
Furthermore, plastic flow is incompressible since, 
\begin{equation}
\textrm{tr} \mathbf{L}^{\textrm{p}} = 0.
\end{equation}

The elastic deformation gradient allows for the polar decomposition,
\begin{equation}
\mathbf{F}^{\textrm{e}}=\mathbf{R}^{\textrm{e}}\mathbf{U}^{\textrm{e}},\\
\end{equation}
where $\mathbf{R}^{\textrm{e}}$ is the elastic rotation and $\mathbf{U}^{\textrm{e}}$ is the elastic right stretch. Then we define the elastic strain tensor as,
\begin{equation}
\begin{aligned}
& \mathbf{E}^{\textrm{e}}=\frac{1}{2}(\mathbf{C}^{\textrm{e}} - \mathbf{1}),\\
\end{aligned}
\end{equation}
where $\mathbf{C}^{\textrm{e}}=\mathbf{F}^{\textrm{e}\textrm{T}}\mathbf{F}^{\textrm{e}}$ is the elastic right Cauchy-Green tensor. Thus, the elastic strain tensor is defined in the intermediate space. 

\subsubsection{Constitutive equations}
The elastic free energy in the intermediate space is defined by,
\begin{equation}
\Phi = \Phi(\mathbf{E}^{\textrm{e}}, \theta) = \frac{1}{2}\mathbf{E}^{\textrm{e}}:\bm{\mathcal{C}}\,[\mathbf{E}^{\textrm{e}}] - (\theta - \theta_0)\mathbf{A} : \bm{\mathcal{C}}[\mathbf{E}^{\textrm{e}}].\\
\end{equation}
Here, `` : '' denotes the inner product of two tensors. Then, the elastic 2nd Piola stress conjugate to $\mathbf{E}^{\textrm{e}}$ is obtained by,
\begin{equation}
\mathbf{T}^{\textrm{e}}= \frac{\partial \Phi(\mathbf{E}^{\textrm{e}}, \theta)}{\partial \mathbf{E}^{\textrm{e}}} =\bm{\mathcal{C}}\,[\mathbf{E}^{\textrm{e}} - \mathbf{A}(\theta - \theta_0)], \\
\end{equation}
where $\bm{\mathcal{C}}$ is the fourth order elastic stiffness tensor and $\mathbf{A}$ is the second order thermal expansion tensor. Moreover, $\theta$ is the current absolute temperature and $\theta_0$ is the reference temperature.
The elastic 2nd Piola stress is related to the Piola stress $\mathbf{P}$ by, 
\begin{equation}
\mathbf{P}=\mathbf{F}^{\textrm{e}}\mathbf{T}^{\textrm{e}}\mathbf{F}^{\textrm{p}-\textrm{T}}.
\end{equation}
Then, it is also related to the Cauchy stress $\mathbf{T}$, using the relation, $\mathbf{T} = J^{-1} \mathbf{P} \mathbf{F}^{\textrm{T}}$ with $J = \mathrm{det} \mathbf{F}$,
\begin{equation}
\label{eqn:cauchy-2pk}
\mathbf{T}^{\textrm{e}} = J \mathbf{F}^{\textrm{e}-\textrm{1}}\textbf{T}\mathbf{F}^{\textrm{e}-\textrm{T}}.
\end{equation}

The resolved shear stress that drives slip on the $\alpha$ th slip system is projected from the elastic 2nd Piola stress via the Schmid tensor defined in the intermediate lattice space,
\begin{equation}
\tau^{\alpha} = \mathbf{C}^{\textrm{e}}\mathbf{T}^{\textrm{e}}:\mathbf{\mathbb{S}}^{\alpha}_0 \approx \mathbf{T}^{\textrm{e}}:\mathbf{\mathbb{S}}^{\alpha}_0,\\
\end{equation}
since the elastic deformation is small in this work.

The plastic strain rate in the $\alpha$ th slip system is then constitutively prescribed by the form of the thermally-activated velocity of screw dislocations,
\begin{equation}
\label{eqn:slip}
\begin{aligned}
\dot{\gamma_\textrm{p}}^{\alpha}=\dot{\gamma_0} \, \textrm{exp} \bigg(-\frac{\Delta G}{k_B \theta}\bigg< 1- \Big( \frac{\tau^{\alpha}_{eff}}{\widetilde{s_l}} \Big)^p \bigg>^q \bigg) \quad \textrm{for} \, \tau^{\alpha}_{eff} > 0 \, \textrm{,} &\\ \textrm{otherwise} \quad \dot{\gamma_\textrm{p}}^{\alpha}=0 \; ; \qquad \tau^{\alpha}_{eff} = \lvert \tau^{\alpha} \rvert - s^{\alpha} &,\\
\end{aligned}
\end{equation}
where $\tau^{\alpha}_{eff}$ is the effective shear stress, $\dot{\gamma_0}$ is the reference slip rate, $\Delta G$ is the activation energy, $k_B$ is Boltzmann’s constant, $s^{\alpha}$ is the slip resistance from dislocation interaction, $p$ and $q$ denote the parameters for the shape of stress-dependent kink-pair formation energy, and $\langle \, \cdot \, \rangle = \frac{1}{2} \big( \, \lvert \,\cdot\,\rvert + (\,\cdot\,)\,\big)$ is a Macaulay bracket. Furthermore, the temperature-dependent lattice resistance , $\widetilde{s_l}$ is expressed by,
\begin{equation}
\widetilde{s_l} = s_l \frac{\mu}{\mu_0},
\end{equation}
where $s_l$ is the lattice resistance at 0 K and $\mu_0=\sqrt{\mathcal{C}_{44,0} \Big( \frac{\mathcal{C}_{11,0}-\mathcal{C}_{12,0}}{2} \Big)}$ and $\mu=\sqrt{\mathcal{C}_{44} \Big( \frac{\mathcal{C}_{11}-\mathcal{C}_{12}}{2} \Big)}$ are the effective shear moduli at 0K and current temperature, respectively. 

\subsubsection{Slip resistance and dislocation evolution}
In classical crystal plasticity theories (\cite{Asaro1983, Asaro1985, Kali, Cuitino1993}), the slip resistance $s^{\alpha}$ in each of the slip systems was taken to evolve according to a simple phenomenological model\footnote[1]{
$\dot{s}^{\alpha} = \sum_{\beta} h^{\alpha \beta} \left| \dot{\gamma}^{\beta} \right|$, where $h^{\alpha \beta}$ is the hardening matrix.} that represents a first description of self- and latent-hardening in the slip systems. This simple phenomenological model was found to reasonably capture the strain-hardening behavior in bcc materials in various crystallographic orientations (\cite{Yalcinkaya2008, Cho2018}). More recently, many single crystal plasticity models for both fcc and bcc materials have employed a modified Taylor hardening law associated with the evolution of dislocation densities (\cite{Bronkhorst2019, Lim2020, Nguyen2021}). In this work, we employ the modified Taylor hardening law to better represent the underlying physics for the evolution of dislocations and their interactions throughout the slip systems. The slip resistance in the $\alpha$-th slip system is expressed as,
\begin{equation}
s^{\alpha} = s_0 + \mu b \sqrt{\sum_{\beta=1}^{N} a^{\alpha\beta} \rho^{\beta}},
\label{eqn:taylor-hardening}
\end{equation}
where $s_0$ is the far-field resistance to slip, $\mu$ is the effective shear modulus, $b$ is the magnitude of Burgers vector and $\rho^{\beta}$ is the dislocation density in each slip system, $\beta$. Moreover, $a^{\alpha\beta}$ is the interaction matrix that characterizes the interaction strength between the slip systems $\alpha$ and $\beta$, which will be further discussed below. 

In order to compute the slip resistance in the modified Taylor hardening law (Equation (\ref{eqn:taylor-hardening})), the dislocation density in each of the slip systems must be simultaneously computed via an appropriate evolution model. Numerous evolution models for computing the dislocation density exist in the literature for single crystal plasticity theories of bcc materials. Some bcc single crystal plasticity models have used a simple evolution rule for dislocation densities in which the interaction strengths between the slip systems were assumed to be equal (\cite{Knezevic2014, Lim2015b}), following the classical work by \cite{Kocks1976}. Moreover, there is vast literature on the bcc single crystal plasticity models in which the geometric features in dislocation interactions throughout the slip systems were taken into account. \cite{Ma2004} and \cite{Ma2007} proposed a simple model in which the forest and parallel dislocation densities associated with the slip resistance in a slip system ($\alpha$) were computed such that dislocation densities in all other slip systems ($\beta \neq \alpha$) were geometrically projected onto the central slip system ($\alpha$). Their interpretation for geometry of dislocation interaction was recently employed to model single crystalline bcc tungsten (\cite{Cereceda2016}) exhibiting anomalous yield features due to non-Schmid effects. More recently, \cite{Nguyen2021} modified the geometric interaction model of Ma, Roters and Raabe to account for evolution of forest and co-planar (including parallel) dislocation densities in which the mixed characteristics of edge and screw dislocations were taken into account. All of these models have found success in capturing some important features in the thermomechanical behavior of single crystal bcc materials in various crystallographic orientations. Yet, these models for bcc single crystal plasticity have limitation in accounting for the dislocation density evolution associated with interaction strengths strongly dependent on the type of interaction as well as the dependence of the dislocation interaction strengths on the hardening behavior. Single crystal plasticity model for bcc materials needs to be further extended in order to more accurately account for the hardening behavior associated with the dislocation microstructure, interaction and generation throughout the slip systems.

Meanwhile, mesoscopic dislocation dynamics (dd) simulations have enabled computing the interaction properties of dislocations for which both short-range contact interactions and long-range elastic interactions are taken into account throughout the slip systems for single crystalline fcc (\cite{Madec2003, Devincre2006}) and bcc materials (\cite{Queyreau2009, Madec2017}). Specifically, for both fcc and bcc crystals, the dd simulations have elucidated the roles of collinear interactions and various junctions in the slip activation processes as well as the hardening mechanisms throughout slip systems, as well postulated in \cite{Madec2003} and \cite{Devincre2005}. Hence, a “soft” multi-scaling between the mesoscopic dd simulation and the macroscopic hardening law has enriched the physical picture in the continuum single crystal models in which the dislocation interaction strengths are explicitly taken into account (\cite{Dequiedt2015, Bronkhorst2019}). In this work, we employ dd simulation results for tantalum to better represent the dislocation microstructures and interactions in our single crystal model.
 
The dislocation density in each of the slip systems in Equation (\ref{eqn:taylor-hardening}) is taken to evolve according to a multiplication-annihilation type model (\cite{Dequiedt2015}),
\begin{equation}
\dot{\rho}^{\; \alpha} = \frac{1}{b} \Bigg(\; \frac{1}{\mathcal{L}^{\alpha}} - 2y_c^{\alpha} \,\rho^{\,\alpha} \;\Bigg) \lvert\dot{\gamma_\textrm{p}}^{\alpha}\rvert 
\end{equation}
where $\mathcal{L}^{\alpha}$ is the mean free path of dislocations, and $y_c^{\alpha}$ is the annihilation capture radius. The mean free path is inversely proportional to the forest dislocation density, i.e.,
\begin{equation}
\label{mean_free}
\frac{1}{\mathcal{L}^{\alpha}} = \sqrt{\sum_{\beta=1}^N d^{\,\alpha\beta} \rho^{\,\beta}}\\
\end{equation}
with $d^{\,\alpha\beta} = \frac{a^{\,\alpha\beta}}{k_1}$ for self interaction or coplanar interaction, and $d^{\,\alpha\beta} = \frac{a^{\,\alpha\beta}}{k_2}$ for other interactions, where $k_1$ and $k_2$ are the mean free path coefficients. In this work, we employ the interaction strengths, $a^{\,\alpha\beta}$, informed by dd simulations for tantalum recently performed by \cite{Madec2017}. Further detailed description of the dislocation interaction strengths employed in our model is provided together with other material parameters in Section \ref{section:para}. Moreover, the temperature and rate-dependent annihilation capture radius is expressed by,
\begin{equation}
y_c^{\alpha} = y_{c0} \Bigg( 1 - \frac{k_B \theta}{A_{rec}} \textrm{ln} \Big\lvert \frac{\dot{\gamma_\textrm{p}}^{\alpha}}{\dot{\gamma}_0}\Big\rvert \Bigg),
\end{equation}
where $y_{c0}$ is the reference annihilation capture radius, and $A_{rec}$ is the capture radius energy, following \cite{Beyerlein2008}.

\subsubsection{Temperature evolution}

It has been known that during plastic deformation of metallic materials, the plastic work is partitioned into stored energy of cold work and thermal energy (\cite{bever1973stored, farren1925heat, taylor1934latent, titchener1958stored}). The energy stored in the atomic bond extension and contraction due to the evolution of dislocation density and structure is significant. The proportion of plastic work partitioned into thermal energy is generally termed the Taylor-Quinney factor (\cite{taylor1934latent}). Although it has been demonstrated that the Taylor-Quinney factor is very likely not constant and can take values substantially below 1.0 (\cite{dorogoy2017dynamic, lieou2020thermodynamic, lieou2021thermomechanical, rittel2012dynamically, rittel2017dependence}), doing so within the present structural theory is beyond the scope of this work and therefore is assumed to simply remain constant. The evolution of temperature is then taken as
\begin{equation}
\label{eqn:temperature}
\rho c  \dot{\theta}= \eta \sum_{\alpha=1}^{N} \tau^{\alpha} \dot{\gamma_\textrm{p}}^{\alpha}, \\
\end{equation}
where $\eta$ is the Taylor-Quinney factor, $\rho$ is the material mass density, and $c$ is the specific heat. Furthermore, it is assumed for strain rates below 1000 s$^{-1}$ that $\eta = 0.0$ and above that $\eta = 1.0$. The Laplacian term in the original heat equation has also been neglected.
\\
\clearpage
\subsubsection{Slip systems and material parameters}
\label{section:para}

As noted in the introduction,
\begin{table}[b!]
\centering
\renewcommand{\arraystretch}{1.2}
\caption{Slip systems for $\{110\} \langle 111 \rangle$}
\begin{tabular}{ m{3cm}  m{2cm} m{2cm}}
\hline
Slip system                & $\mathbf{m}^{\alpha}_0$  & $\mathbf{{n}}^{\alpha}_0$ \\
\hline
A2                 &[$\overline{1}$11]   &  (0$\overline{1}$1)\\
A3                 &[$\overline{1}$11]   &  (101)\\
A6                 &[$\overline{1}$11]   &  (110)\\
B2                 &[111]  &  (0$\overline{1}$1)\\
B4                 &[111]  &  ($\overline{1}$01)\\
B5                 &[111]   & ($\overline{1}$10) \\
C1                 &[$\overline{1} \overline{1}$1]  & (011) \\
C3                 &[$\overline{1} \overline{1}$1]  & (101) \\
C5                 &[$\overline{1} \overline{1}$1]  & ($\overline{1}$10)\\
D1                 &[1$\overline{1}$1]   & (011) \\
D4                 &[1$\overline{1}$1] &  ($\overline{1}$01)\\
D6                 &[1$\overline{1}$1] &  (110)\\
\hline
\end{tabular}
\label{table:slip_110}
\end{table}
\begin{table}[b!]
\centering
\renewcommand{\arraystretch}{1.2}
\caption{Slip systems for $\{112\} \langle 111 \rangle$}
\begin{tabular}{ m{3cm}  m{2cm} m{2cm}}
\hline
Slip system                & $\mathbf{m}^{\alpha}_0$  & $\mathbf{{n}}^{\alpha}_0$ \\
\hline
A$\underline{4}$                 &[$\overline{1}$11]   &  (211)\\
A$\underline{8}$                 &[$\overline{1}$11]   &  ($\overline{1}$1$\overline{2}$)\\
A$\underline{11}$                 &[$\overline{1}$11]   &  ($\overline{1} \overline{2}$1)\\
B$\underline{3}$                 &[111]  &  ($\overline{2}$11)\\
B$\underline{7}$                 &[111]  &  (11$\overline{2}$)\\
B$\underline{12}$                 &[111]   & (1$\overline{2}$1) \\
C$\underline{2}$                 &[$\overline{1} \overline{1}$1]  & ($\overline{1}\overline{1}\overline{2}$)) \\
C$\underline{5}$                 &[$\overline{1} \overline{1}$1]  & ($\overline{1}$21) \\
C$\underline{10}$                 &[$\overline{1} \overline{1}$1]  & (2$\overline{1}$1)\\
D$\underline{1}$                 &[1$\overline{1}$1]   & (1$\overline{1}\overline{2}$) \\
D$\underline{6}$                 &[1$\overline{1}$1] &  (121)\\
D$\underline{9}$                 &[1$\overline{1}$1] &  ($\overline{2}\overline{1}$1)\\
\hline
\end{tabular}
\label{table:slip_112}
\end{table}
screw dislocations in tantalum and other bcc metallic materials dissociate into non-planar partial dislocation configurations while at rest as a lower energy state. This creates ambiguity in the proper stress to use to drive dislocation motion as the partial dislocation configuration is believed to be composed of three Burgers vectors which form a triangle. To provoke motion of screw dislocations, the split core must be forced to become planar once again and given the triangular configuration of the partial dislocations, the stress conditions to do so are not directionally isotropic and set up a condition of twin and anti-twin directionality for the motion of these dislocations on any given slip system. Atomistic calculations of screw dislocation dissociation in a number of different bcc materials on the $\{110\} \langle 111 \rangle$ type of systems has been clearly demonstrated (\cite{Duesbery1973, Groger2008a, Groger2008b, Yang2001}). Similar demonstration for the $\{112\} \langle 111 \rangle$ and  $\{123\} \langle 111 \rangle$ types has not yet been made. Neither has the influence of the split dislocation core upon dislocation interactions been studied in bcc metals given the hypothesis that screw dislocation motion may be via the nucleation and propagation of kink-bands (\cite{butler2018mechanisms}). We know from prior work that including both $\{110\} \langle 111 \rangle$ and $\{112\} \langle 111 \rangle$ types of slip systems as options for dislocation motion is necessary to properly describe crystallographic texture evolution(\cite{Bronkhorst2006,Kothari1998}) with the role of $\{123\} \langle 111 \rangle$ systems within a continuum crystal mechanics setting unclear. Although, not yet well quantified, there are indications that the directional asymmetry described above is reduced with increase in material temperature (\cite{Lim2015a}). Prior work has estimated this to be noticeable but small at room temperature for tantalum (\cite{Bronkhorst2021, Cho2018}). Therefore, given the current uncertainties presented, we employ the traditional Schmid tensor and corresponding resolved shear stress as the primary external driving force for dislocation motion on the $\{110\} \langle 111 \rangle$ and $\{112\} \langle 111 \rangle$ types and treat all slip systems equally. Slip systems are listed in Table \ref{table:slip_110} and \ref{table:slip_112}.

The material parameters used in the model are listed in Table \ref{table:para}.
\begin{table}[b!]
\centering
\renewcommand{\arraystretch}{1.3}
\caption{Material parameters used in this study}
\begin{tabular}{l@{\quad}l @{\qquad \quad} l@{\quad}l}
\hline

$\rho$ [kg/m$^3$]                & 16640  & $a_{\textrm{J}}$               & 0.05 \\
$c$  [J/kg-K ]                 & 150    & $a_{\textrm{XJ}}$              & 0.04 \\
$\alpha$  [$\mu$m/m-K]           & 6.5    & $s_0$ [MPa]                    & 35.0 \\
$k_b$ [J/K]                      & 1.38 $\times$ 10$^{-23}$  & $\dot{\gamma_0}$ [sec$^{-1}$]  & 1.0 $\times$ 10$^{7}$ \\
$C_{11, 0}$ [GPa]                & 268.5  & $\Delta G$ [J]               & 2.1 $\times$ 10$^{-19}$       \\
$C_{12, 0}$ [GPa]                & 159.9  & $s_l$ [MPa]                    & 400.0 \\
$C_{44, 0}$ [GPa]                & 87.1   & $\sum_{\alpha} \rho^\alpha_{0}$ [m$^{-2}$] & 2.4 $\times$ 10$^{12}$ \\
$m_{11}$ [MPa/K]                 & -24.5  & $y_{c0}$                       & $6b$ \\
$m_{12}$ [MPa/K]                 & -11.8  & $A_{rec}$ [J]                  & 2.0 $\times$ 10$^{-20}$ \\
$m_{44}$ [MPa/K]                 & -14.9  & $p$                            & 0.28 \\
$a_{\textrm{copl}}$              & 0.06   & $q$                            & 1.34 \\
$a_{\textrm{colli}\,60\degree}$   & 0.7744 & $k_1$                          & 180 \\
$a_{\textrm{colli}\,90\degree}$   & 0.9025 & $k_2$                          & 2.5 \\
$a_{\textrm{colli}\,30\degree}$   & 0.5112 & $b$ [nm]                       & 0.286 \\
\hline
\label{table:para}
\end{tabular}
\end{table}
The mass density $\rho$, specific heat $c$, and thermal expansion tensor $\mathbf{A} = \alpha \mathbf{1}$ are assumed to be constant during deformation. The fourth order elastic stiffness tensor is expresed in terms of the three independent elastic constants ($\mathcal{C}_{11}$, $\mathcal{C}_{12}$, $\mathcal{C}_{44}$). The elastic constants are linearly dependent on temperature as $\mathcal{C}_{ij} = \mathcal{C}_{ij,0} + m_{ij} \theta$. $\{  \mathcal{C}_{11,0}, \, \mathcal{C}_{12,0}, \, \mathcal{C}_{44,0} \}$ at 0 K and $\{ m_{11}, \, m_{22}, \, m_{33} \}$ are assumed to follow the previous work by \cite{Kothari1998}, \cite{Cho2018} and \cite{Bronkhorst2021}. The values of $\dot{\gamma_0}$, $\Delta G$, $s_l$ and $p$ and $q$ in the flow model are identified based on the previous work that employed similar single crystal models for tantalum (\cite{Kothari1998, Cho2018, Bronkhorst2021}). Then, these values have been further tuned to better capture the rate-dependent stress-strain behaviors in the single crystals.

The dislocation interaction coefficients are taken to follow the work by \cite{Madec2017}. Instead of using the full asymmetric interaction matrices throughout the slip systems of $\{$110$\} \langle$111$\rangle$ and $\{$112$\} \langle$111$\rangle$, we further simplified the interaction coefficients, as follows.\\
\noindent
$\bullet$ \quad $a_{\textrm{copl}}$ = 0.06 \quad for the self- and coplanar interactions (\cite{Dequiedt2021}),\\
$\bullet$ \quad $a_{\textrm{colli}\,60\degree}$ = 0.7744 \quad for the collinear interaction and $\theta = \arccos \lvert \mathbf{n}^{\alpha}_0 \cdot \mathbf{n}^{\beta}_0 \rvert = 60 \degree$,\\
$\bullet$ \quad $a_{\textrm{colli}\,90\degree}$ = 0.9025 \quad for the collinear interaction and $\theta = \arccos \lvert \mathbf{n}^{\alpha}_0 \cdot \mathbf{n}^{\beta}_0 \rvert = 90 \degree$,\\
$\bullet$ \quad $a_{\textrm{colli}\,30\degree}$ = 0.5112 \quad for the collinear interaction and $\theta = \arccos \lvert \mathbf{n}^{\alpha}_0 \cdot \mathbf{n}^{\beta}_0 \rvert = 30 \degree$,\\
$\bullet$ \quad $a_{\textrm{J}}$ = 0.05 \quad for the junctions between $\{$110$\}$ systems or $\{$112$\}$ systems,\\
$\bullet$ \quad $a_{\textrm{XJ}}$ = 0.04 \quad for the junctions between $\{$110$\}$ and $\{$112$\}$ systems.\\
As pointed out in \cite{Monnet2006}, the presence of the friction stress due to alloy friction or lattice resistance can screen the elastic field between dislocations. Thus, this culminates in a decrease in the line-tension and interaction strengths. Since the lattice resistance has been taken into account in our single crystal model, the values for the junction strengths ($a_{\textrm{J}}, a_{\textrm{XJ}}$) are taken to be smaller than the average junction strengths determined in \cite{Madec2017}.

The values for the remaining parameters, $s_0, \, y_{c0},\, A_{rec}, \, k_1, \, \textrm{and} \, k_2$ are then identified using the experimental data for the single crystal behaviors at low to high strain rates and for the polycrystal behavior at low strain rate in through-thickness direction. 

\subsection{Results: experiment vs. model}

The finite deformation single crystal model presented in Section \ref{section:single_plasticity} was numerically implemented for use in a finite element solver (a standard branch of Abaqus) for boundary value problems of single- and polycrystalline tantalum discussed in the next sections. We implemented the implicit multi-step computational procedure for updating the stress tensor, kinematic tensors and all of the state variables including the dislocation densities evolving together with deformation, following and modifying the algorithms proposed by  \cite{Kali} and \cite{Anand2004}. Furthermore, we computed a forth order tangent tensor also known as the Jacobian consistent with the viscoplastic single crystal model used in a Newton-type iteration for obtaining a solution that satisfies the global equilibrium at the end of the increment in the nonlinear boundary value problems, based on the procedures detailed in  \cite{dai1997} and \cite{balasubramanian1998}. All the details regarding the implicit time integration procedure and the computation of tangent are provided in the Appendices \ref{appendix:Time_int} and \ref{appendix:Jac}, respectively.

Here, we validate the finite deformation single crystal model presented above against experimental data for single crystal tantalum. We used the single crystal stress-strain data at low to high strain rates recently published by \cite{Nguyen2021}, \cite{Rittel2009} and \cite{Whiteman2019}.

\subsubsection{Single crystal behavior at low to high strain rate}

\begin{figure}[h!]
	\centering
    \includegraphics[width=1\textwidth]{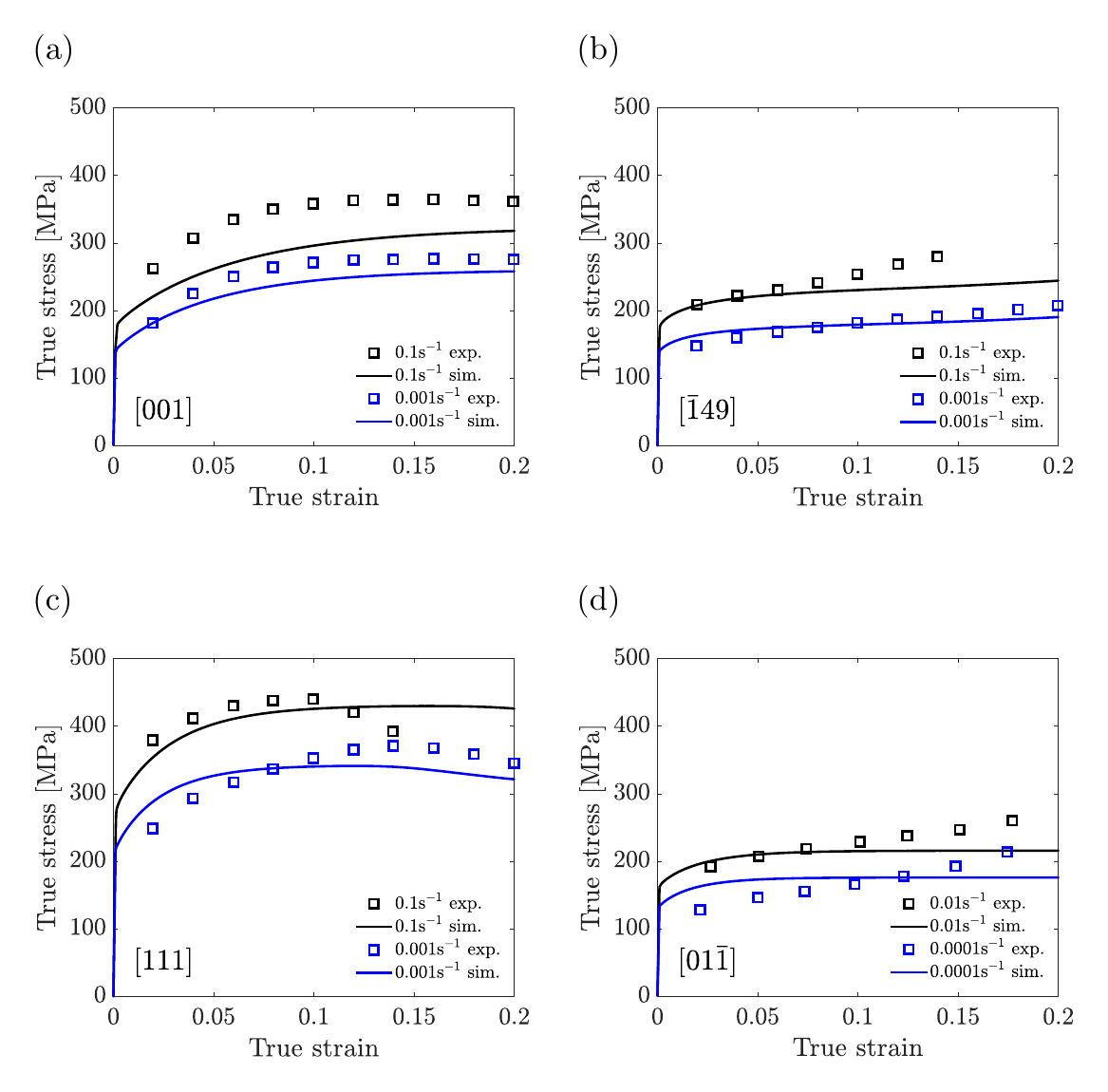}
\caption{Stress-strain behavior of single crystal tanlaum in experiments and numerical simulations at low strain rates: (a) [001], (b)[$\bar{1}$49], (c)[111] and (d)[01$\bar{1}$] in compression. The single crystal data were taken from \cite{Lim2020} and \cite{Rittel2009}.}
\label{fig:result1}
\end{figure}

Figure \ref{fig:result1} shows the measured and numerically simulated stress-strain curves for single crystalline tantalum in crystallographic orientations of [001], [01$\bar{1}$], [111] and [$\bar{1}$49] at strain rates of 0.001 s$^{-1}$ and 0.1 s$^{-1}$. As evidenced in the figures, the model captures well the main features of the rate-dependent yield and flow stresses for the various crystallographic directions. The highest initial yield point and the flow stresses in the [111] orientation due to the lowest Schmid factor are nicely described in the model. Furthermore, the hardening behavior and its tendency towards larger strains are reasonably captured in the [001], [111] and [$\bar{1}$49] orientations. However, in the [01$\bar{1}$] orientation, the hardening behavior observed in the experiment is poorly predicted by the model. This discrepancy can likely be attributed to crystallographic misorientation\footnote[2]{As pointed out by \cite{Cuitino1993} and \cite{Stainier2002}, for single crystal specimens with high symmetric orientation along the loading direction, small misalignment in the loading axis breaks the symmetry of resolved shear stress, resulting in a significant change in slip activity and stress responses.} or some finite size effects in samples during experimentation. Additionally, the stress-strain behavior in this particular [01$\bar{1}$] orientation has large variations amongst the experimental data reported by different research groups (\cite{Rittel2009, Lim2020, Whiteman2019}).

\begin{figure}[h!]
\centering
    \includegraphics[width=1\textwidth]{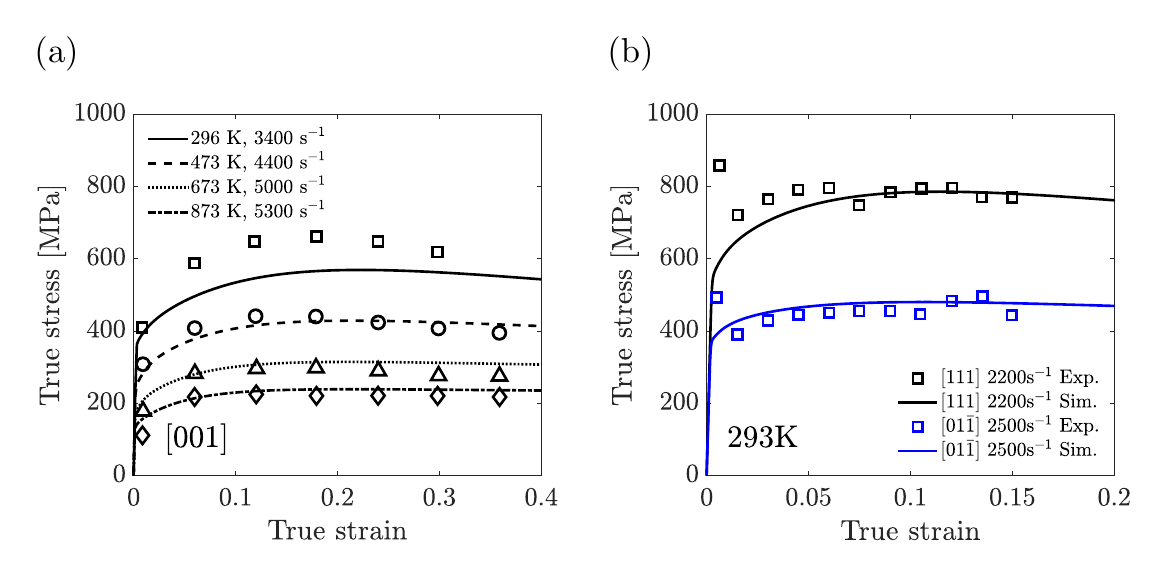}
\caption{Stress-strain behavior of single crystal tantalum in experiments (open symbols) and numerical simulations (lines) at high strain rate compression. (a) [001] at room temperature to higher (b) [01$\bar{1}$] and [111] at high strain rates. The single crystal data were taken from \cite{Nguyen2021} and \cite{Whiteman2019}.}
\label{fig:result2}
\end{figure}

In Figure \ref{fig:result2}, the single crystal behavior is further presented at high strain rates in both experiments and numerical simulations. The high strain rate curves are displayed for room temperature and higher in the crystallographic orientations of [001] (Figure \ref{fig:result2}a), [01$\bar{1}$] and [111] (Figure \ref{fig:result2}b). The orientation- and temperature-dependent yield stress and flow stresses are excellently captured by the model. Furthermore, the single crystal model captures nicely the thermal softening that manifests due to adiabatic heating under high strain rate conditions. In summary, our single crystal model has been shown to be capable of capturing the main features of the orientation-dependent stress-strain behaviors of single crystal tantalum at low (0.001 s$^{-1}$) to high ($\sim$ 5000 s$^{-1}$) strain rates and at room (296K) to high (873K) temperatures.

\subsubsection{Effects of collinear interactions on slip activity and instability in single crystals}

The collinear interactions taken into account in our single crystal model have been found to be of critical importance for predicting active slip systems in single crystals loaded in orientations with high symmetry (e.g. $\langle$100$\rangle$ and $\langle$110$\rangle$), as posited by \cite{Madec2003} and \cite{Devincre2005}. \cite{Devincre2005} also reported on important experimental results for fcc single crystals loaded in the $\langle$100$\rangle$ orientation. In their experiment, though eight primary slip systems had the same Schmid factor, only four of them were found to be activated under deformation, attributed to instability induced by the strong collinear interactions between the eight slip systems; i.e., only one in a collinear pair of two slip systems was found to be activated for slip. Here, we analyze the role of collinear interactions in the slip activation processes for this exemplar tantalum single crystal. We also numerically examine instability induced by strong collinear interactions, especially on $\{$110$\}$ slip planes. Towards this end, we excluded the 12 $\{$112$\}$ $\langle$111$\rangle$ slip systems in the numerical simulations for single crystals \footnote[3]{If the $\{$112$\}$ slip systems are included, no slip systems in collinear interactions are activated in single crystal samples loaded in the high symmetry orientations of $\langle$100$\rangle$, $\langle$110$\rangle$ and $\langle$111$\rangle$. Hence, no instability is observed as shown in our results in Figure \ref{fig:result1} and \ref{fig:result2}.}. 

\begin{figure}[h!]
	\centering
    \includegraphics[width=1\textwidth]{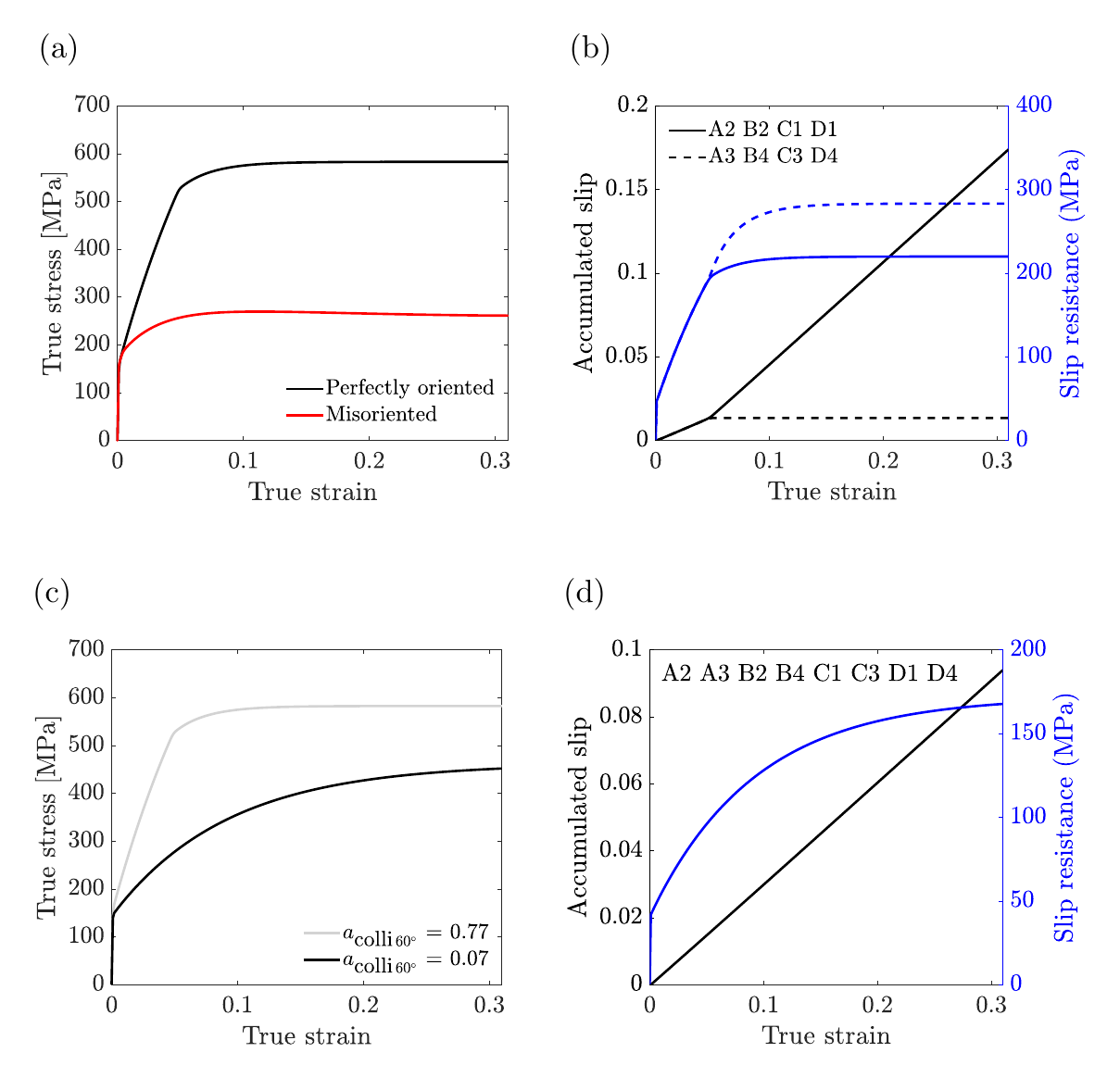}
\caption{Effect of collinear interactions on single crystal behavior loaded in high symmetry orientations. (a) Stress-strain curves with and without misorientation, (b) evolution of accumulated slip and slip resistance without misorientation, (c) stress-strain curves and (d) accumulated slip and slip resistance with a weaker collinear interaction.}
\label{fig:colli}
\end{figure}

Figure \ref{fig:colli} shows a numerically simulated stress-strain response in a tantalum single crystal loaded in a high symmetry orientation of $\langle$100$\rangle$, together with accumulated slips and slip resistances in eight slip systems having the same Schmid factor. Here, (A3, A2), (B4, B2), (C3, C1) and (D4, D1) are the pairs of slip systems collinear-interacting on the $\{$110$\}$ slip planes. As shown in Figure \ref{fig:colli}b, initially, though the accumulated slips in these eight slip systems evolve at the same rate, they bifurcate after a small amount of deformation. Interestingly, in the deactivated cross-slip systems (A3, B4, C3 and D4), the slip resistance continues to evolve. While, in the active slip systems (A2, B2, C1 and D1), it does not evolve significantly since their cross-slip counterparts become deactivated (See the modified Taylor hardening law in Equation (\ref{eqn:taylor-hardening})). The instability in the slip activation process was found to culminate in an anomalous stiff response beyond the initial yield as shown in Figure \ref{fig:colli}a (black solid line with no misorientation). Such an anomalous stress response sustains until the collinear interactions vanish due to the lattice distortion in the deformed single crystal. Next, we introduced a slight misorientation (0.6 $\degree$) between the sample axis and the loading direction. 
The slight misorientation resulted in asymmetry in the initial Schmid factors (or resolved shear stresses) for the eight slip systems. The asymmetry due to the slight misorientation culminated in early determination for activation throughout the slip systems. Hence, as evidenced in the simulated stress-strain curve (red solid line in Figure \ref{fig:colli}a), the abnormally stiff response observed in the perfectly oriented sample diminishes significantly with misorentation. 

The single crystal response in this high symmetry orientation is further examined with a weaker collinear interaction strength. As shown in Figure \ref{fig:colli}c, the single crystal exhibits a stress-strain behavior without any precursor for instability. Such a stable hardening behavior is clearly evidenced by Figure \ref{fig:colli}d on slip resistances and accumulated slips in the eight slip systems. These slip systems having the same Schmid factor are equally active without any bifurcation during deformation. These results strongly support the idea that the collinear interactions play a critical role in determining slip activation processes in bcc single crystals loaded in high symmetry orientations. It should be noted that the slip activation mechanisms as well as the instability phenomena cannot be accounted for by the early bcc single crystal models with the phenomenological hardening law or the modified Taylor hardening law without any notion of collinear interactions informed by the mesoscopic dislocation dynamics simulations. It should also be noted that the instability induced by the strong collinear interactions is particularly important in analyzing the bcc single crystal behavior where slips on the $\{$110$\}$ planes are more likely to be activated (e.g. \cite{Cho2018, Lim2015b, Narayanan2014, Patra2014}).

\section{Polycrystal behavior}
\label{section:polycrystal}

The deformation mechanisms in polycrystal tantalum have been investigated, traced back to the early work by \cite{Hoge1977}, \cite{Chen1996}, \cite{Nemat-Nasser1998} and \cite{Kothari1998}. In these classical papers, the highly temperature- and rate-dependent inelastic features in polycrystal tantalum were addressed. Furthermore, the polycrystal models that employed the phenomenological hardening law was found to capture well some features in flow stresses and crystallographic texturing without any notion on dislocation evolution and interaction. This success with the phenomenological hardening law has been deemed due to the macroscopic hardening responses not sensitive to the details of dislocation evolution and interaction throughout the polycrystalline network.

Here, the single crystal model presented in the Section \ref{section:single_cry} is further underpinned by examining the predictive capabilities of the model for polycrystal tantalum. To this end, we conducted mechanical tests for polycrystal tantalum samples machined in two different directions (through-thickness and in-plane) taken from the wrought tantalum plate. Furthermore, \textit{ex situ} measurements on crystallographic texture evolution in the deformed samples at increasing strains were conducted. In addition to crystallographic texture, dislocation density was monitored using \textit{ex situ} neutron diffraction measurements. The polycrystalline behavior involving stress-strain responses, dislocation density evolution and texture evolution is then reproduced in numerical simulations in which the single crystal plasticity model is employed without any further modification. The results below show the predictive capabilities of our modeling framework on both single- and polycrystal tantalum materials at low to high strain rate and at room temperature and higher.

\subsection{Experiment}
Multiple cylindrical tantalum specimens of 4.2 mm diameter and 8.4 mm length were electro-discharge machined from a wrought plate with their axes parallel to either the through-thickness (TT) or in-plane (IP) directions. IP polycrystal tantalum specimens were compressed in-situ at a strain rate of 0.001 s$^{-1}$ to maximum true strain levels of 0.08, 0.18, 0.44, 0.89 or 0.149. In parallel, the TT and IP specimens were compressed \textit{ex situ} to compressive true strains of 0.2, 0.3 and 0.4.

Bulk texture measurements of the specimens were conducted on the High-Pressure/Preferred Orientation (HIPPO) neutron time-of-flight diffractometer (\cite{Wenk2003, Vogel2004}) at the Lujan Center at the Los Alamos Neutron Science Center (LANSCE). HIPPO consists of 1,200 $^3$He detector tubes on 45 panels arranged on five rings around the incident neutron beam with nominal diffraction angles of 144$\degree$, 120$\degree$, 90$\degree$, 60$\degree$, and 40$\degree$ covering 22.4$\%$ of 4$\pi$ (\cite{Takajo2018}). The deformed samples at strains of 0.2, 0.3 and 0.4 were glued on to sample holders with their cylinder axis along the holder axis and loaded on an automated robotic sample changer on HIPPO (\cite{Losko2014}). Data was then collected at three rotations around the sample axis of 0$\degree$, 67.5$\degree$, and 90$\degree$. Data over a d-spacing range from 0.7Å to 2.5Å was analyzed with the Rietveld method (\cite{Rietveld1969}) as implemented in the Materials Analysis Using Diffraction (MAUD) code following procedures described previously (\cite{Wenk2010}). The orientation distribution was represented by the E-WIMV method (\cite{Matthies2005}) using a resolution of 7.5$\degree$. From the MAUD analysis, pole figure data recalculated from the refined ODF was exported for further processing.

High-resolution (FWHM ∼0.1$\%$), high-statistics time-of-flight (TOF) neutron diffraction data was collected on the Spectrometer for MAterials Research at Temperature and Stress (SMARTS) for the purpose of diffraction line profile analysis (DLPA) (\cite{Brown2016}) of each of the deformed specimens. The extended convolutional multiple whole profile (eCMWP) method (\cite{Ribarik2004}) was used for semi-quantitative determination of the dislocation density in each of the deformed tantalum specimens. Annealed copper foil was used to determine the instrumental resolution and a Pearson \uppercase\expandafter{\romannumeral7} function was used to fit the individual peak profiles to determine the breadth and shape parameters. 

\subsection{Results: experiment vs. numerical simulation}
\label{section:poly_result}

Numerical simulations for polycrystalline tantalum are conducted in which each of the finite elements represents one crystal. We use a finite element model comprising 1000 (10$\times$10$\times$10; Figure \ref{fig:result4}a) hexahedral elements with reduced integration. For simple compression, all faces are constrained to remain parallel, and displacement boundary conditions corresponding to loading conditions are applied on the top surface. The initial texture in each of the through-thickness and the in-plane direction was extracted from the measured data using MTEX (\cite{Bachmann2010}) and the corresponding set of Euler angles ($\phi$, $\theta$ and $\omega$) was then randomly assigned to each of the elements in order to represent the initial material texture.

\subsubsection{Stress-strain behavior, crystallographic texturing and dislocation density evolution}
\label{section:poly}
\begin{figure}[h!]
\centering
    \includegraphics[width=1\textwidth]{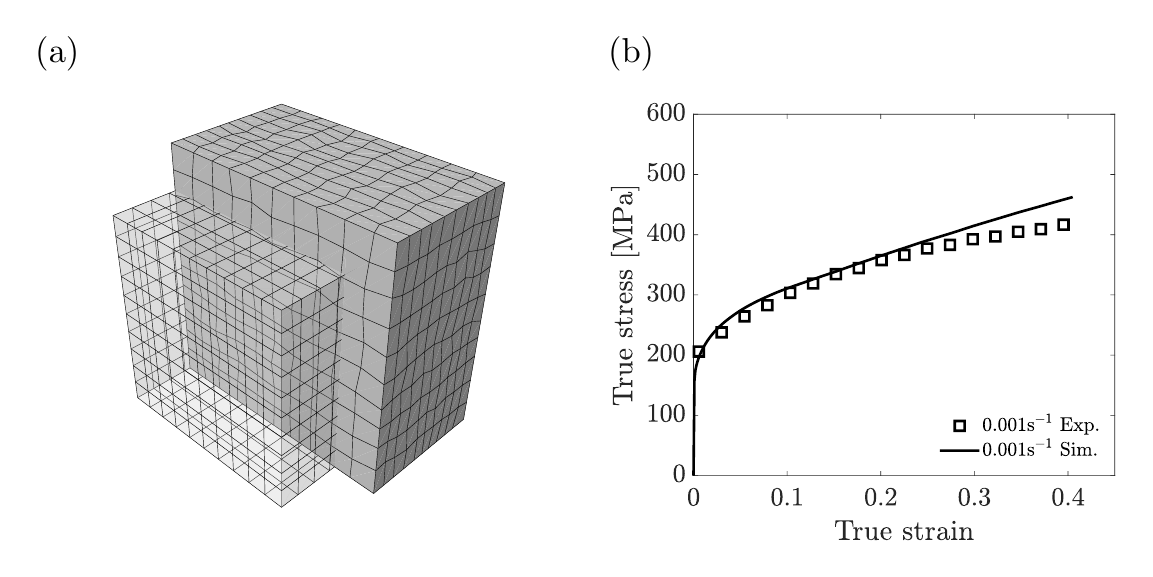}
\caption{(a) polycrystal model in undeformed (transparent) and deformed (gray) configurations, (b) stress-strain response in experiment and numerical simulation under loaded along the through-thickness direction at 0.001 s$^{-1}$ and 300K.}
\label{fig:result4}
\end{figure}

Figure \ref{fig:result4} shows the response of polycrystal tantalum in compression (strain rate: 0.001 s$^{-1}$) in both experiment and numerical simulation. Together with undeformed and deformed meshes shown in Figure ~\ref{fig:result4}a, the measured and numerically simulated stress strain curves along the through-thickness direction are displayed in Figure \ref{fig:result4}b. As shown, the numerically simulated responses including yield stress and overall hardening behavior are in good agreement with the measured data.

\begin{figure}[h!]
\centering
\hspace*{0.1in}
\includegraphics[width=1\textwidth]{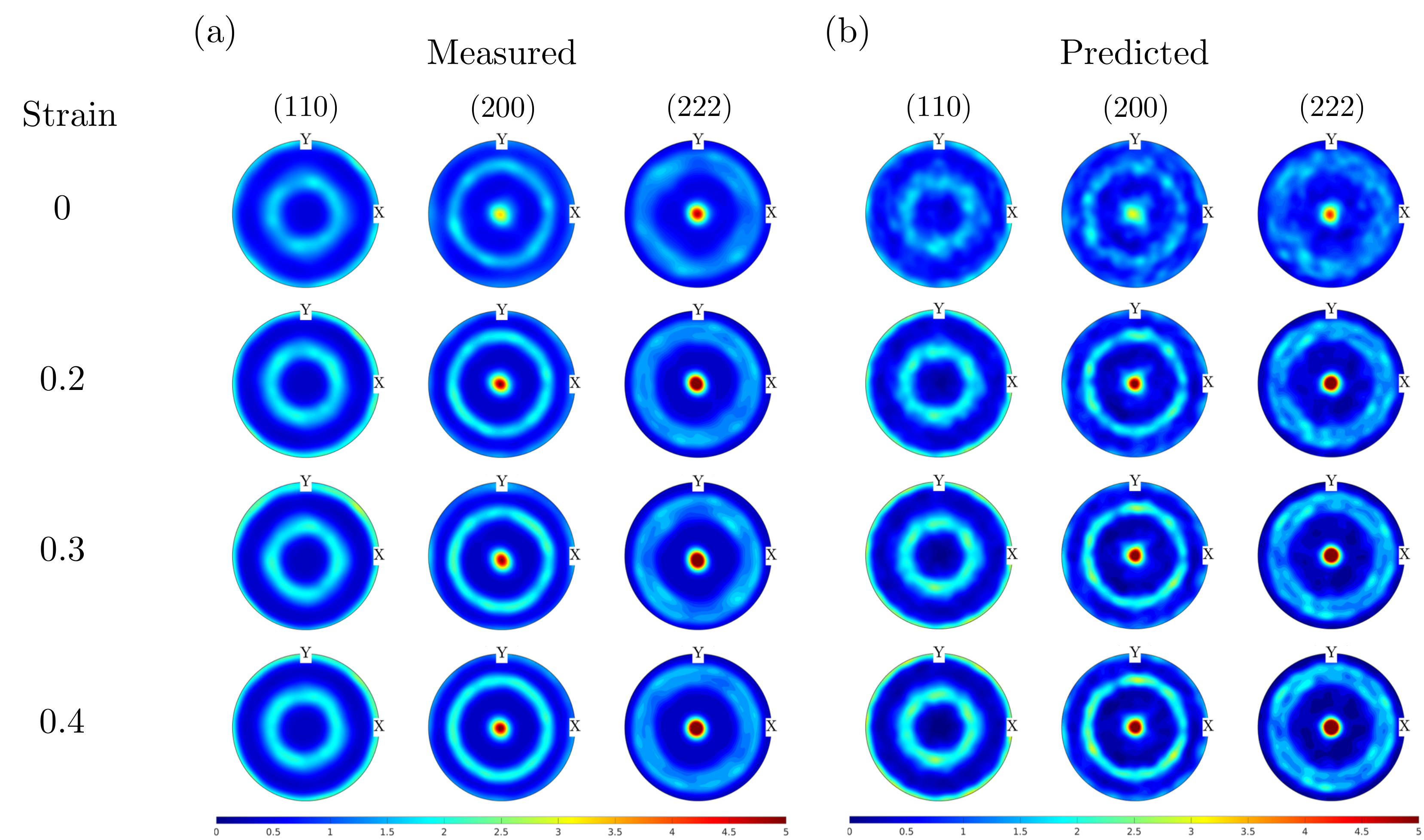} 
\caption{Texture evolution during compression along the through-thickness direction. Pole figures (a) measured and (b) numerically predicted.}
\label{fig:pole_tt}
\end{figure}

\begin{figure}[h!]
\centering
\hspace*{0.1in}
\includegraphics[width=1\textwidth]{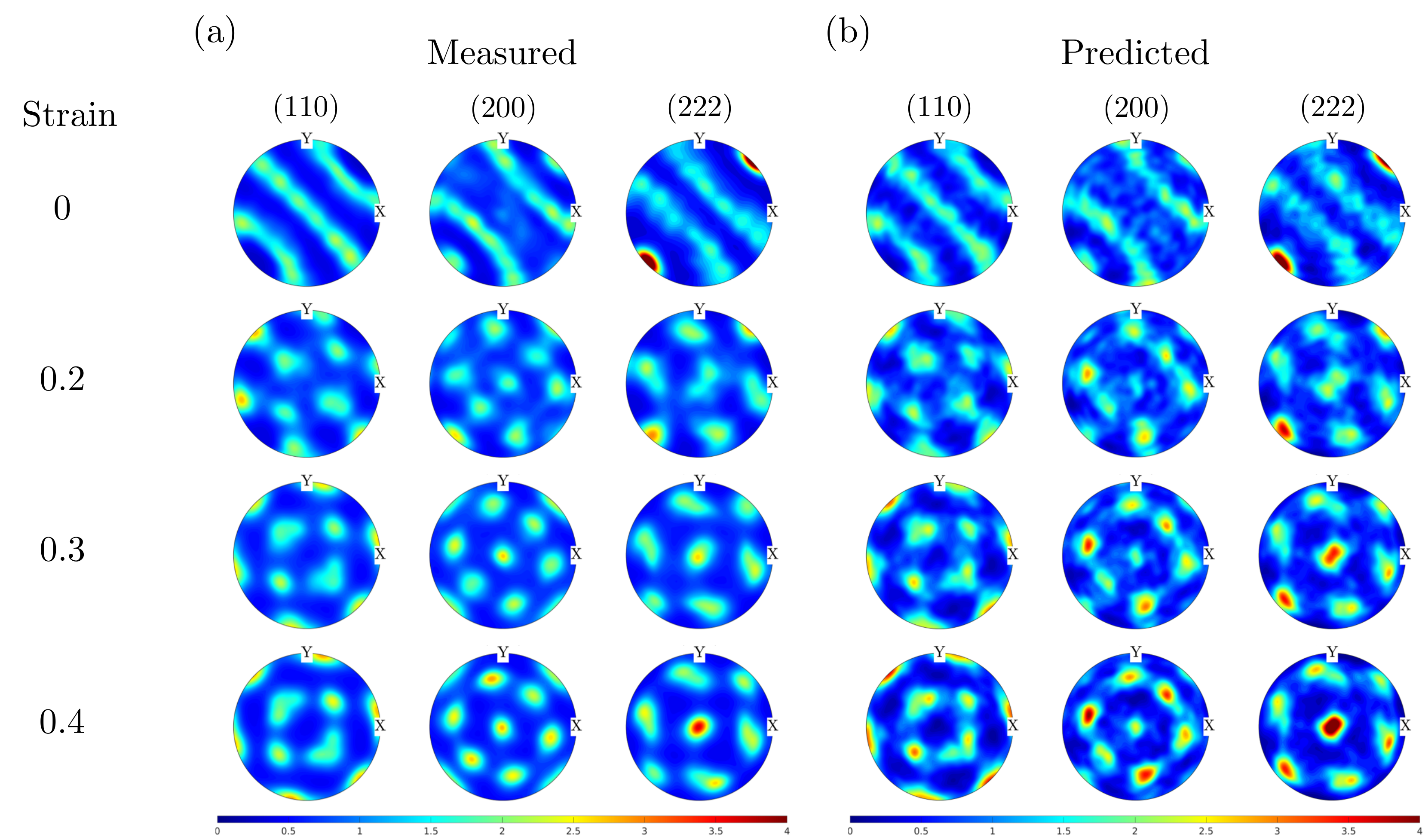} 
\caption{Texture evolution during compression along the in-plane direction. Pole figures (a) measured and (b) numerically predicted.}
\label{fig:pole_ip}
\end{figure}

Figures \ref{fig:pole_tt} and \ref{fig:pole_ip} shows a comparison of the pole figures from experiments and numerical simulations of the polycrystal tantalum under compression along the through-thickness and in-plane directions respectively. The sample in the through-thickness direction was found to be initially textured in the (100) and (111) as shown in Figure \ref{fig:pole_tt}a at zero strain. In addition, the sample in the in-plane direction was found to be textured in the (110) direction as shown in Figure \ref{fig:pole_ip}a at zero strain. The texture development toward the (100) and (111) directions is attributed to constrained rotations of grains by prescribed slip systems (\cite{kocks1998texture}), excellently predicted in our numerical simulations for both sample directions. Moreover, as recently pointed out in \cite{Lim2020} the rotation characteristics in bcc crystals are strongly dependent upon the types of the prescribed slip systems. Our experimental and numerical results on the texture evolution strongly support that the prescribed slip systems on both $\{$110$\}$ and $\{$112$\}$ planes have been appropriately chosen for this study.

\begin{figure}[h!]
\centering
\includegraphics[width=1\textwidth]{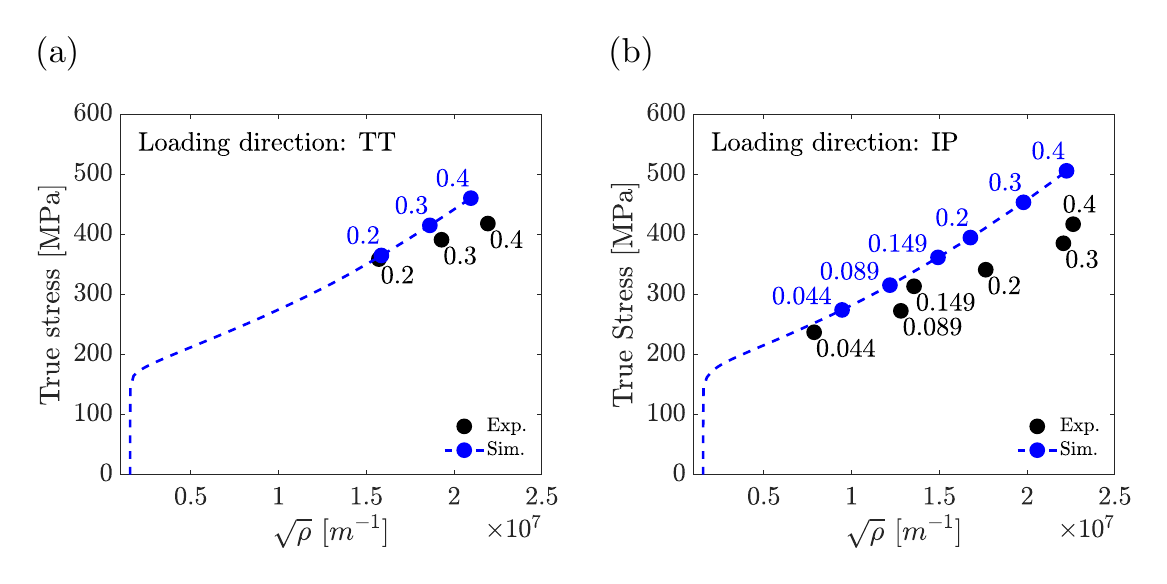} 
\caption{Comparison between dislocation densities measured and numerically predicted for (a) through-thickness direction and (b) in-plane direction.}
\label{fig:result_dd}
\end{figure}

We then directly compared the dislocation densities measured in our \textit{ex situ} neutron diffraction experiments to those predicted in our numerical simulations in the through-thickness direction (Figure \ref{fig:result_dd}a) and the in-plane direction (Figure \ref{fig:result_dd}b). To this end, we plotted the flow stresses in the two sample directions as functions of dislocation density and imposed strain. Here, since the volume change is small, the total dislocation density is simply calculated by averaging the dislocation density at each integration point. As shown, our numerical simulations capture reasonably the relations between dislocation densities and macroscopic stress and strain responses in both sample directions in terms of trend. Furthermore, the flow stresses are linearly dependent on the square root of the averaged dislocation density in both experiments and numerical simulations. However, the overall stress response in the numerical simulation especially in the in-plane direction is found to be higher than that in the experiment. This discrepancy is presumably attributed to a lack of information on initial distributions of dislocation densities for numerical simulations and not yet adequately representing the complexities of dislocation dynamics and intergranular interaction in the materials.


\subsubsection{High strain rate and high temperature behavior}

\begin{figure}[h!]
	\centering
\includegraphics[width=1\textwidth]{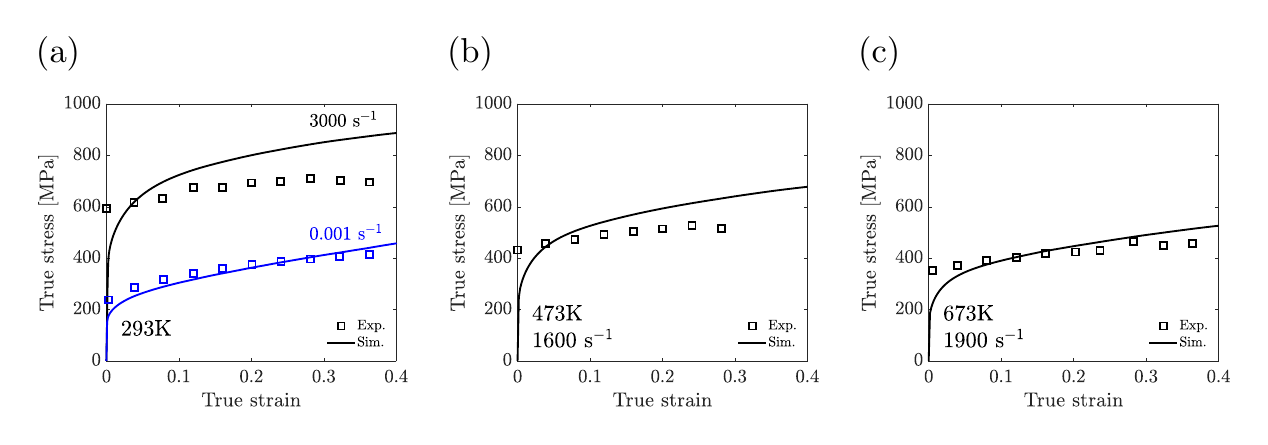}
\caption{Stress-strain behavior of polycrystal tanlaum in experiments and numerical simulations at high strain rates and high temperatures: (a) at 298K 0.001 s$^{-1}$ and 3000 s$^{-1}$, (b)at 473K 1600 s$^{-1}$, and (c) at 673K 1900 s$^{-1}$ in compression.}
\label{fig:poly_high}
\end{figure}

Figure \ref{fig:poly_high} shows the stress-strain behavior of polycrystal tantalum in compression at higher temperature(T $\geq$ 298K) and high strain rate ($\dot{\epsilon} > 10^3$ s$^{-1}$) in both experiments and numerical simulations. Here, the high strain rate data have been collected using a split Hopkinson pressure bar system. As shown in Figure \ref{fig:poly_high}a, the numerical simulations predict the low- to high strain rate behavior in polycrystal tantalum at room temperature reasonably well. Furthermore, the remarkable decreases in yield stress and flow stresses at higher temperatures of 473 K and 673 K are well captured in the numerical simulation displayed in Figure \ref{fig:poly_high}b and \ref{fig:poly_high}c. Although the overall high strain rate- and high temperature behavior is reasonably predicted, the high strain rate hardening features at room temperature and 473 K are poorly captured in the numerical simulations. It is presumably attributed to inaccuracy in the interaction properties computed from dd simulations especially at large strains and at high strain rates.

\
\subsection{Grain-level analysis for rotation and dislocation density evolution}

The macroscopic features in dislocation density, crystallographic texturing and stress-strain responses in polycrystal tantalum samples have been found to be captured well by the simple polycrystal model in which each element represented one crystal. Here, the polycrystal behavior is further addressed by numerical simulations on a polycrystal model based on Voronoi-tessellation. Though this Voronoi-tessellation-based polycrystal model is not grain-boundary conforming, it enables analysis of the local variations of dislocation density, slip activity, stress and rotation throughout more realistic polycrystalline network not available in the simple polycrystal model presented in Section \ref{section:poly_result}.

\begin{figure}[h!]
\centering
\includegraphics[width=1\textwidth]{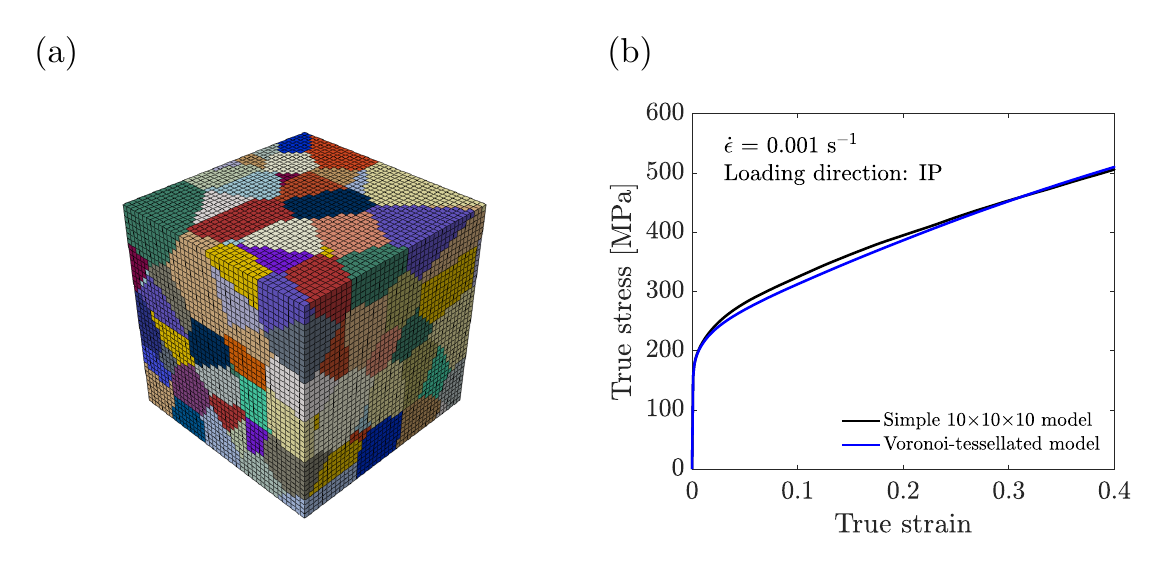}
\caption{(a) Voronoi-tessellation-based polycrystal model with 200 random spatial points (b) comparison of stress-strain curves between 10 $\times$ 10 $\times$ 10 polycrystal model and Voronoi model in the in-plane direction.}
\label{fig:result_tess}
\end{figure}

Figure \ref{fig:result_tess}a shows a Voronoi-tessellated unit-cube with 200 random spatial Voronoi points. The tessellated domain is meshed with the hexahedral elements via voxelization using an open-source program, Neper (\cite{Quey2011}). Moreover, the number of random Voronoi points has been chosen such that the Voronoi model reproduces the texture evolution captured by the simple polycrystal model (the set of 10 $\times$ 10 $\times$ 10 hexahedral elements) displayed in Figure \ref{fig:pole_ip}. In the Voronoi-tessellation-based polycrystal model, each of the polyhedral Voronoi cells represents one crystal to which a set of Euler angles extracted from the experiment is assigned. Then we conducted numerical simulations for the polycrystal domains under large compression. As shown in Figure \ref{fig:result_tess}b, the macroscopic stress-strain curve in the Voronoi-tessellation-based polycrystal model matches well with that for the simple polycrystal model in the in-plane direction. 

\begin{figure}[h!]
	\centering
\includegraphics[width=1\textwidth]{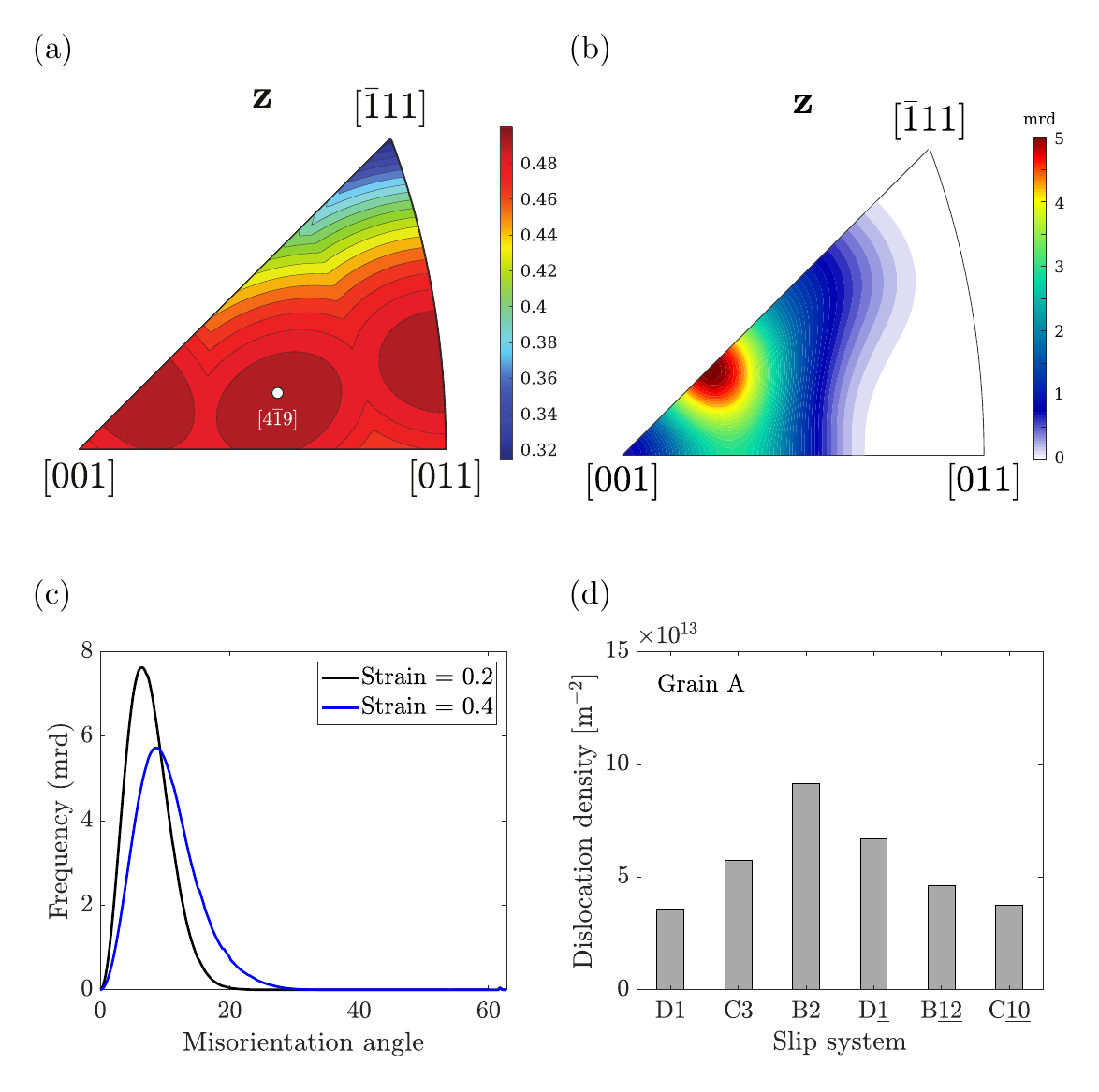}
\caption{Numerically predicted rotation and dislocation density in a grain A in the [4$\bar{1}$9] direction within the Voronoi-tessellated polycrystal domain. (a) Location of the [4$\bar{1}$9] direction on contour of the maximum Schmid factor, (b) countour of crystallographic orientations within the grain A at a strain of 0.4, (c) distribution of misorentation angle within the grain (strains of 0.2 and 0.4), (d) dislocation densities in slip systems on both $\{$110$\}$ and $\{$112$\}$ planes at a strain of 0.4.}
\label{fig:result_g3}
\end{figure}

In the Voronoi-tessellation-based polycrystal model, three grains have been selected for detailed analysis for a intragranular behavior. The grain A initially oriented in the [4$\bar{1}$9] direction was selected since the location in the standard triangle has the maximum Schmid factor on the $\{$110$\}$ slip planes (Figure  \ref{fig:result_g3}a). Under deformation, as shown in Figure \ref{fig:result_g3}b, the grain tends to rotate toward the edge (between [001] and [1$\bar{1}$1] of the standard triangle) and then toward [001] such that the Schmid factor of the initially active slip system decreases along the corresponding great circle (\cite{dieter1976mechanical}). Furthermore, a significant variation in rotation (or intragranular orientation) is observed inside the grain, as shown in Figures \ref{fig:result_g3}b and \ref{fig:result_g3}c. Herein, the intragranular misorientation is computed via, $\Delta \mathbf{g}_i = \mathbf{g}_i \langle \mathbf{g}_{\textrm{avg}} \rangle^{-1}$, where $\Delta \mathbf{g}_i$ is the magnitude of misorientation between the orientation of a spatial intragranular point $i$ ($\mathbf{g}_i$) and the average orientation in the grain ($\mathbf{g}_{\textrm{avg}}$), following \cite{Pokharel2014}. Together with the intragranular rotation map, dislocation densities in major active slip systems at a strain of 0.4 are displayed in Figure \ref{fig:result_g3}d. Though the Schmid factor on the $\{$110$\}$ slip systems (e.g. the slip system of B2) is initially greater than on the $\{$112$\}$ slip planes, the dislocation densities are found to develop well on both $\{$110$\}$ and $\{$112$\}$ planes, due to the remarkable spatial variation in rotation inside the grain; i.e., at a strain of 0.4 (Figure \ref{fig:result_g3}c), the contour for local orientation in the grain A is widely located throughout the regions where the Schmid factors are strong on both $\{$110$\}$ and $\{$112$\}$ slip planes.

\begin{figure}[h!]
	\centering
\includegraphics[width=1\textwidth]{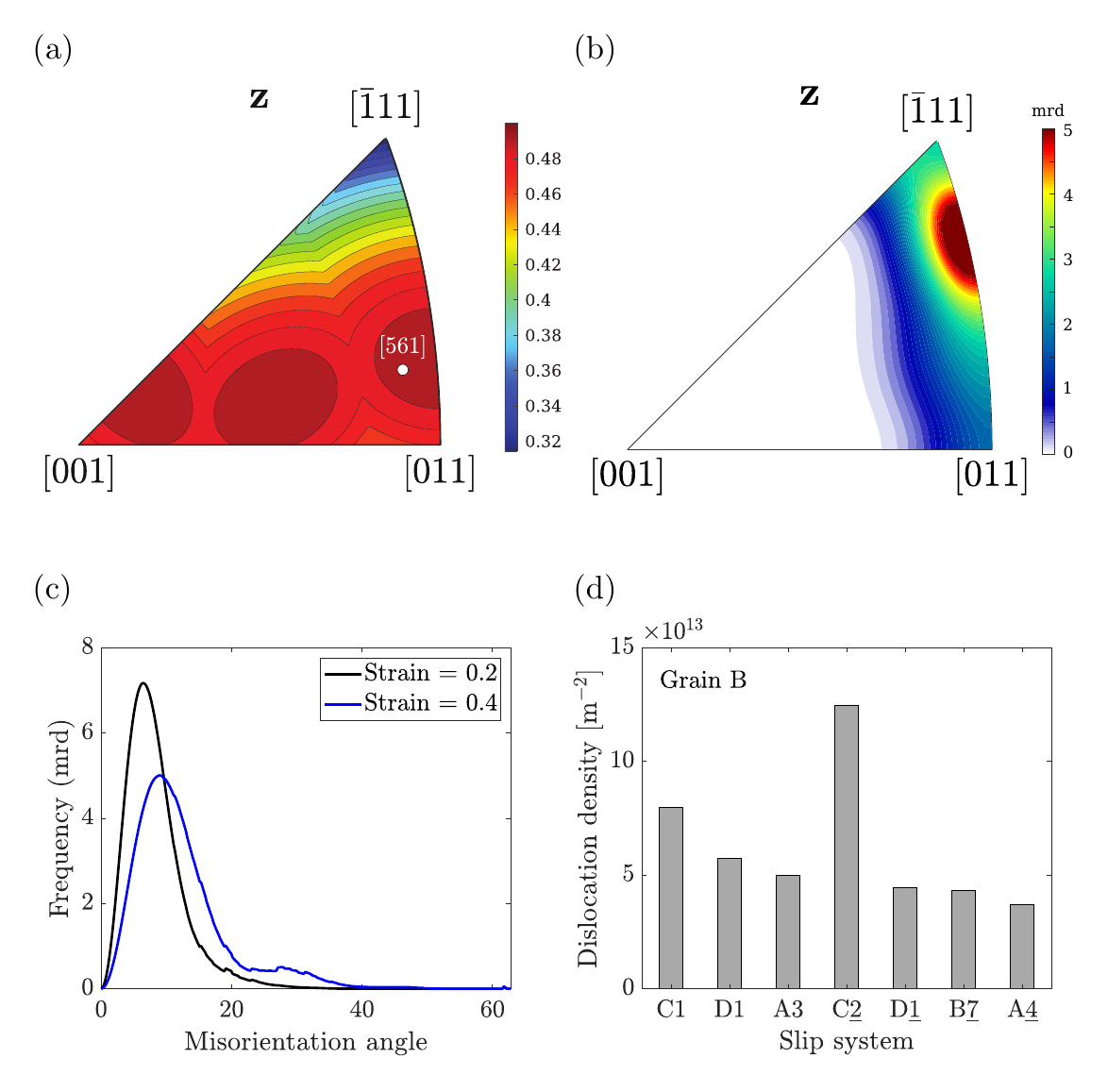}
\caption{Numerically predicted rotation and dislocation density in a grain B in the [561] direction within the Voronoi-tessellated polycrystal domain. (a) Location of the [561] direction on contour of the maximum Schmid factor, (b) countour of crystallographic orientations within the grain B at a strain of 0.4, (c) distribution of misorentation angle within the grain (strains of 0.2 and 0.4), (d) dislocation densities in slip systems on both $\{$110$\}$ and $\{$112$\}$ planes at a strain of 0.4.}
\label{fig:result_g30}
\end{figure}

Then, a grain B initially oriented in the [561] was selected since the initial orientation in the standard triangle has the maximum Schmid factor on the $\{$112$\}$ planes (Figure \ref{fig:result_g30}a). As shown in Figure \ref{fig:result_g30}b, the grain tends to rotate toward the [111] direction such that the Schmid factor decreases along the corresponding great circle. Moreover, once again a significant spatial variation in rotation is observed again in the grain B at increasing strains of 0.2 and 0.4, as displayed in Figure \ref{fig:result_g30}c. Dislocation densities in major active slip systems at a strain of 0.4 are also shown in Figure \ref{fig:result_g30}d. The dislocation density is found to grow remarkably, especially for a slip system on the $\{$112$\}$ plane (here, C$\underline{2}$). This is quite reasonable since the contour for local orientation in the grain B is still located in the region where the maximum Schmid factor is incurred in the slip systems on the $\{$112$\}$ planes. 

\begin{figure}[h!]
	\centering
\includegraphics[width=1\textwidth]{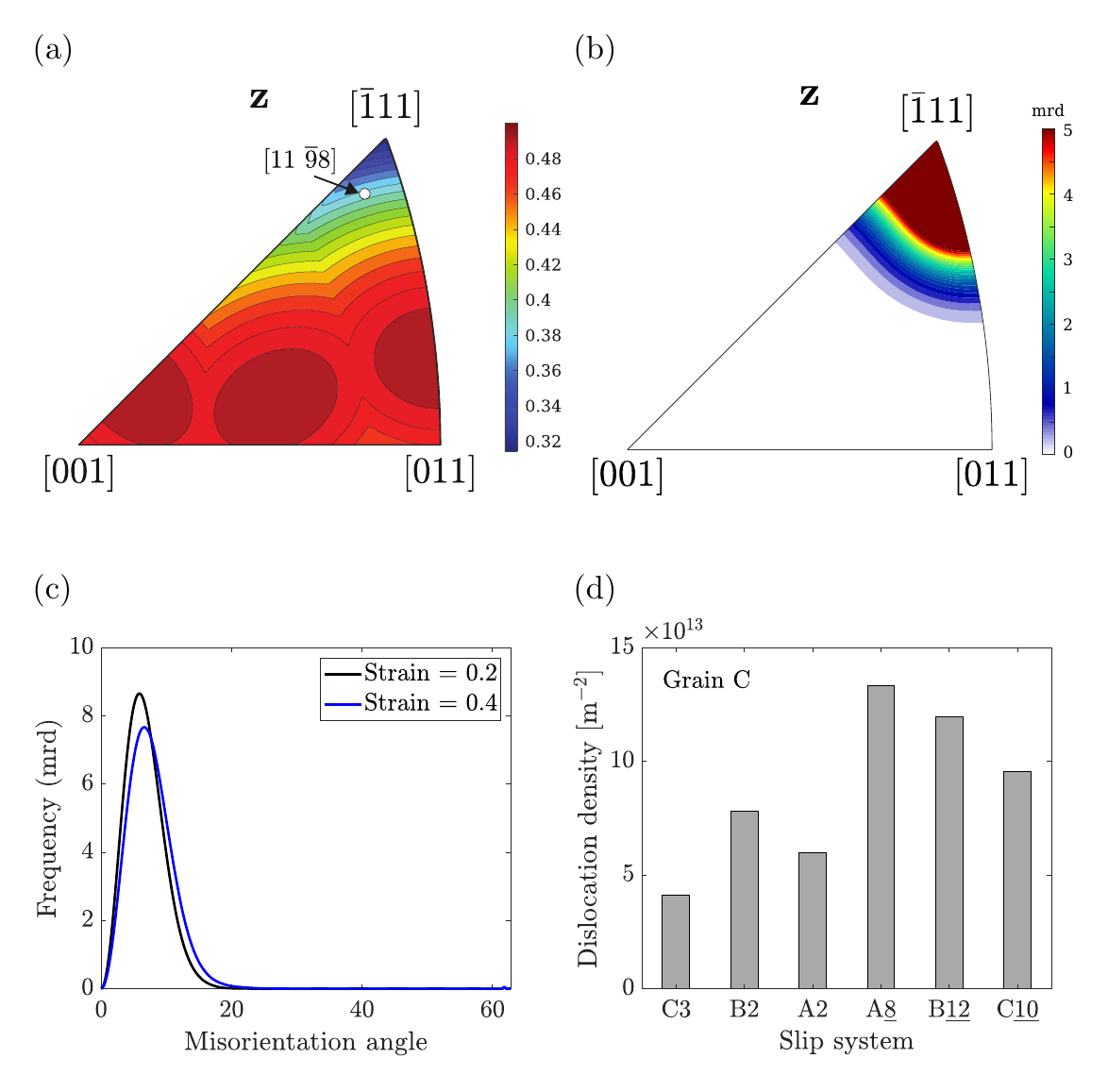}
\caption{Numerically predicted rotation and dislocation density in a grain C in the [11\,$\bar{9}$8] direction within the Voronoi-tessellated polycrystal domain. (a) Location of the [11\,$\bar{9}$8] direction on contour of the maximum Schmid factor, (b) countour of crystallographic orientations within the grain C at a strain of 0.4, (c) distribution of misorentation angle within the grain (strains of 0.2 and 0.4), (d) dislocation densities in slip systems on both $\{$110$\}$ and $\{$112$\}$ planes at a strain of 0.4.}
\label{fig:result_g116}
\end{figure}

Lastly, Figure \ref{fig:result_g116} shows the intragranular behavior of a grain C initially oriented in the 
[11\,$\bar{9}$8] direction. The grain C was selected since it was closely aligned to [1$\bar{1}$1] direction (Figure \ref{fig:result_g116}a). It rotates much less than the other two grains presented in Figure \ref{fig:result_g3} and \ref{fig:result_g30}. Furthermore, as shown in Figure \ref{fig:result_g116}b and \ref{fig:result_g116}c, there is no significant intragranular misorientation. Dislocation densities in the grain C grow very similarly to those in a single crystal loaded in the [1$\bar{1}$1] direction, attributed to the small rotation during deformation. As shown in Figure \ref{fig:result_g116}d, the dislocation densities in the grain C grow remarkably in the slip systems of A\underline{8}, B\underline{12} and C\underline{10} that exhibit the highest Schmid factors in a single crystal loaded in the [1$\bar{1}$1] direction.

\clearpage
\section{Discussion and concluding remarks}
\label{section:discussion}

Plastic deformation in bcc tantalum has been deemed a complex process which involves various length-scales. The inelastic behaviors in bcc tantalum are strongly rate- and temperature-dependent, involving complex physics in the motion of dominant screw dislocations. In past decades, continuum mechanics-based crystal plasticity theories have been proposed to address the salient features of the rate-and temperature-dependent deformation in this important refractory metallic material. Furthermore, the continuum crystal plasticity theories have enabled modeling of the large inelastic behavior of tantalum single crystals or polycrystals. In this work, we have extended the crystal plasticity model to elucidate the deformations of both single- and polycrystal tantalum materials at a wide range of strain rates and temperatures. Furthermore, using a suite of finite deformation constitutive modeling, experiments and numerical simulations, we have addressed the main features of deformation mechanisms, dislocation evolution and the hardening behavior in bcc tantalum.

At the single crystal level, we extended the finite deformation single crystal viscoplasticity model to account for the evolution and interaction of dislocation densities throughout the slip systems on both $\{$110$\}$ and $\{$112$\}$ planes. Our single crystal model takes into account the dislocation interaction strengths critically associated with the hardening behavior as well as the microstructural evolution in bcc single crystals, informed by the recent dislocation dynamics simulation results for tantalum. The single crystal model with material parameters simply calibrated with experimental data on single crystal tantalum has been shown to accurately capture the major features in the stress-strain responses at low to high strain rates and at room temperature and higher in the major crystallographic orientations of [001], [01$\bar{1}$], [111] and [$\bar{1}$49]. Furthermore, using the single crystal model, we addressed the effects of collinear interactions on slip activity and instability in single crystals loaded in high symmetry orientations. The strong collinear interactions were found to give rise to bifurcation in the slip activity throughout the slip systems. The instability under high symmetry loading directions was also found to diminish with a slight misorientation. Though our single crystal model successfully captures the key features of the large inelastic behavior of single crystal tantalum at 0.001 s$^{-1}$ to 5300 s$^{-1}$ and at room temperature and higher, it needs to be further developed to account for low temperature behavior accompanied by the non-Schmid effects, especially on the $\{$110$\}$ $\langle$111$\rangle$ slip systems. To this ends, the interaction properties throughout the slip systems must be further identified by more improved dislocation dynamics simulations, especially below room temperature. Furthermore, our analysis for the effects of collinear interaction and misorientation on instability in the slip activity should be of critical importance in further developing the model able to capture the single crystal behavior at low temperature, in which slip on the $\{$110$\}$ planes are more likely to be activated.

Our single crystal model has been further underpinned by the successful representation of polycrystal behaviors in both experiments and numerical simulations. We conducted mechanical testing of polycrystal tantalum specimens together with \textit{ex situ} neutron diffraction measurements for crystallographic textures and dislocation density at increasing strains. Numerical simulation results showed predictive capabilities of our modeling framework for the macroscopic stress-strain behavior and the corresponding crystallographic texturing at low to high strain rates and at room temperature and higher. Our polycrystal model was also shown to be capable of reproducing the Taylor’s relation in which the flow stress is linearly proportional to the square root of the dislocation density experimentally evidenced in \textit{ex situ} neutron diffraction measurements on dislocation density growth in the polycrystal specimen at increasing strains. Furthermore, we have investigated spatial variations of rotation and dislocation density throughout the more realistic polycrystal network upon using numerical simulations on a Voronoi-tessellation-based polycrystal network. The numerically predicted local variations of the inelastic features provided critical insight into the underlying physical pictures for microstructural evolution within the individual crystals undergoing severe plastic deformation which recently have received great attention for polycrystal materials (\cite{Pokharel2014, Millett2020, Charpagne2021}).

Our structurally unified modeling framework has been shown not only to reproduce the experimental data but also provide physical insight into the deformation mechanisms and microstructural evolutions in single crystal and polycrystal tantalum under plastic deformation at a broad range of strains, strain rates and temperatures. Though our model captures many important aspects of plastic deformation in bcc tantalum, much needs to be done for further refinement of the theory for bcc materials. In particular, the model as proposed is structural only and therefore does not yet account for the proper partition of energy into evolution of dislocation structure or thermal energy. This is particularly important for the ability of predicting the evolution of dislocation density and structure and properly accounting for the evolution of temperature within a thermally sensitive theory. Furthermore, there is much we do not yet know about the mechanics and energetics of bcc dislocation behavior and interaction within the potential kink-pair mechanism of dislocation motion, particularly for the rate limiting screw dislocations and the role which edge dislocations play (\cite{butler2018mechanisms, wang2021generalized}). Indeed, the bcc crystal plasticity theory needs to be further extended to account for the influence of grain boundaries and thereby grain size effect that critically influence the characteristics of dislocation motion in and out of grains of a polycrystal, consequently culminating in substantial change in the macroscopic inelastic features. In addition, short- or long-range spatial interactions between dislocations and grain boundaries or between different grains throughout the polycrystal network should be taken into account for accurately modeling extreme events such as damage nucleation and growth. Towards this end, the more realistic, grain boundary conforming polycrystal models must be established using the electronic backscatter diffraction and \textit{state-of-the-art} measurement techniques for the actual microstructures for bcc materials (\cite{foster2021towards}).
\section*{Acknowledgement}

This work was funded by National Research Foundation (NRF) of Korea (Grant No. 2020R1-C1C101324812 and 2021R1A4A103278311) and National Science Foundation (NSF) CMMI (Grant No. 2115399). This work has also benefitted from the use of the Los Alamos Neutron Science Center (LANSCE) at Los Alamos National Laboratory. Los Alamos National Laboratory is operated by Triad National Security, LLC, for the National Nuclear Security Administration of the U.S. Department of Energy under contract number 89233218NCA000001.

\begin{appendices}
\section{Time integration procedure for single crystal model}
\label{appendix:Time_int}
\renewcommand{\theequation}{\thesection.\arabic{equation}}
\setcounter{equation}{0}

The implicit time integration procedure we have used in this work is summarized, as follows.
Let $\tau= t + \Delta t$. Using the given,

\begin{enumerate}
\item $\mathbf{F}(t),\, \mathbf{F}(\tau)$,
\item $\mathbf{F}^{\textrm{p}}(t),\, s^{\alpha}(t),\, \mathbf{T}(t)$
\end{enumerate}
our task is to update $\{ \mathbf{F}^{\textrm{p}}(\tau), \, s^{\alpha}(\tau), \, \rho^{\alpha}(\tau) \, \textrm{and} \, \mathbf{T}(\tau) \}$.

Components of the fourth order elasticity tensor in the global basis are calculated by,
\begin{equation}
\mathcal{C}_{ijkl} = Q_{ip} Q_{jq} Q_{kr} Q_{lr} \mathcal{C}^{c}_{pqrs},
\end{equation}
where $\mathcal{C}^{c}_{pqrs}$ is the component of the fourth order elasticity tensor in the crystal basis and $\mathbf{Q}$ is the orthogonal tensor which rotates the crystal basis to the global basis (e.g. \cite{Anand2004}). Trial quantities are then calculated by,
\begin{equation}
\mathbf{F}^{\textrm{e}}_{\textrm{tr}}=\mathbf{F}(\tau)\mathbf{F}^{\textrm{p}-1}(t), \\ 
\end{equation}
\begin{equation}
\mathbf{C}^{\textrm{e}}_{\textrm{tr}}=\mathbf{F}^{\textrm{e} \textrm{T}}_{\textrm{tr}}\mathbf{F}^{\textrm{e}}_{\textrm{tr}}, \\ 
\end{equation}
\begin{equation}
\mathbf{E}^{\textrm{e}}_{\textrm{tr}}=\frac{1}{2}(\mathbf{C}^{\textrm{e}}_{\textrm{tr}} - \mathbf{1}), \\ 
\end{equation}
\begin{equation}
\mathbf{T}^{\textrm{e}}_{\textrm{tr}} = \bm{\mathcal{C}}\,[\mathbf{E}^{\textrm{e}}_{\textrm{tr}} - \mathbf{A}(\theta - \theta_0)], \\
\label{eqn:trial_stress}
\end{equation}
\begin{equation}
\label{eqn:b_alpah}
\mathbf{B}^{\alpha} = \textrm{symm} [ \mathbf{C}^{\textrm{e}}_{\textrm{tr}} \mathbf{\mathbb{S}}^{\alpha}_0 ], \\
\end{equation}
\begin{equation}
\label{eqn:c_alpah}
\mathbf{C}^{\alpha} =\bm{\mathcal{C}}\, \Big[ \frac{1}{2} \mathbf{B}^{\alpha}\Big]. \\
\end{equation}
The following coupled implicit equations for $\mathbf{T}^{\textrm{e}}(\tau),\, s^{\alpha}(\tau)$ and $\rho^{\alpha}(\tau)$,
\begin{equation}
\label{eqn:stress_vs_trial_stress}
\mathbf{T}^{\textrm{e}}(\tau)=\mathbf{T}^{\textrm{e}}_{\textrm{tr}} - \sum_{\alpha=1}^{N} \Delta {\gamma_\textrm{p}}^{\alpha}\big(\mathbf{T}^{\textrm{e}}(\tau), \, s^{\alpha}(\tau) \big) \mathbf{C}^{\alpha},\\
\end{equation}
\begin{equation}
s^{\alpha}(\tau) = s^{\alpha}(t) + \Delta s^{\alpha} \big(\Delta \gamma_\textrm{p}^{\beta}\big(\mathbf{T}^{\textrm{e}}(\tau), \, s^{\beta}(\tau) \big),\, \rho^{\beta}(\tau) \big),\\
\end{equation}
\begin{equation}
\rho^{\alpha}(\tau) = \rho^{\alpha}(t) + \Delta \rho^{\alpha} \big(\Delta \gamma_\textrm{p}^{\alpha}\big(\mathbf{T}^{\textrm{e}}(\tau), \, s^{\alpha}(\tau) \big),\, \rho^{\beta}(\tau) \big),
\end{equation}
with,
\begin{equation}
\label{eqn:n+1_slip}
\begin{aligned}
\Delta \gamma_\textrm{p}^{\alpha}= \Delta t \,  \dot{\gamma_\textrm{0}} \, \textrm{exp} \bigg(-\frac{\Delta G}{k_B \theta}\bigg< 1- \Big( \frac{\lvert \tau^{\alpha} \rvert - s^{\alpha}(\tau)}{\widetilde{s_l}} \Big)^p \bigg>^q \bigg); \quad \textrm{where} \, \tau^{\alpha} = \mathbf{T}^{\textrm{e}}(\tau):\mathbf{\mathbb{S}}^{\alpha}_0,
\end{aligned}
\end{equation}
\begin{equation}
\begin{aligned}
\Delta s^{\alpha} &= \Delta t \, \dot{s}^{\alpha}(\tau)\\ &= \frac{1}{2} \mu \frac{\sum_{\beta} a^{\alpha \beta}}{\sqrt{\sum_{\beta} \; a^{\alpha \beta} \rho^{\beta}(\tau)}} \Bigg(\sqrt{\sum_{\gamma} d^{\beta \gamma} \rho^{\gamma}(\tau)} - 2 y_c^{\beta} \rho^{\beta}(\tau) \Bigg) \big\lvert \Delta \gamma_\textrm{p}^{\beta} \big(\mathbf{T}^{\textrm{e}}(\tau), \, s^{\beta}(\tau) \big) \big\rvert,
\end{aligned}
\end{equation}
\begin{equation}
\Delta \rho^{\alpha} = \Delta t \, \dot{\rho^{\alpha}}(\tau) = \frac{1}{b} \Bigg( \sqrt{\sum_{\beta} d^{\alpha \beta} \rho^{\beta}(\tau)} - 2 y_c^{\alpha} \rho^{\alpha}(\tau) \Bigg)\lvert \Delta \gamma_\textrm{p}^{\alpha}\big(\mathbf{T}^{\textrm{e}}(\tau), \, s^{\alpha}(\tau) \big) \rvert,
\end{equation}
where,
\begin{equation}
y_c^{\alpha} = y_{c0} \Bigg( 1 - \frac{k_B \theta}{A_{rec}} \textrm{ln} \Big\lvert \frac{\Delta \gamma_\textrm{p}^{\alpha}}{\Delta t \, \dot{\gamma}_0}\Big\rvert \Bigg),
\end{equation}
are solved using the following two-step iteration procedure.

First, the elastic 2nd Piola stress is updated, keeping $s^{\alpha}(\tau)$ and $\rho^{\alpha}(\tau)$ fixed, by,
\begin{equation}
\label{eqn:newton_correction}
\mathbf{T}_{n+1}^{\textrm{e}}(\tau) = \mathbf{T}_{n}^{\textrm{e}}(\tau) - \mathrsfso{J}^{-1}_n [\mathbf{G}_{\textrm{n}}],\\
\end{equation}
\begin{equation}
\mathbf{G}_{\textrm{n}} \equiv \mathbf{T}_{n}^{\textrm{e}}(\tau) - \mathbf{T}^{\textrm{e}}_{\textrm{tr}} + \sum_{\alpha=1}^{N} \Delta {\gamma_\textrm{p}}^{\alpha}\big(\mathbf{T}^{\textrm{e}}_{n}(\tau), \, s^{\alpha}_k(\tau) \big) \mathbf{C}^{\alpha},
\end{equation}
\begin{equation}
\mathrsfso{J}_n \equiv \mathbfcal{I} + \sum_{\alpha=1}^{N} \mathbf{C}^{\alpha} \otimes \frac{\partial}{\partial \mathbf{T}_{n}^{\textrm{e}}(\tau)}\Delta {\gamma_\textrm{p}}^{\alpha} \big(\mathbf{T}^{\textrm{e}}_{n}(\tau), \, s^{\alpha}_k(\tau) \big),
\end{equation}
where $\mathbfcal{I}$ is the fourth-order identity tensor. The elastic 2nd Piola stress is accepted if,
\begin{equation}
\Big\lvert \big[ \mathrsfso{J}^{-1}_n [\mathbf{G}_{\textrm{n}}] \big]_{ij} \Big\rvert < \Delta T^{\textrm{e}}_{\textrm{tol}},
\end{equation}
where $\Delta T^{\textrm{e}}_{\textrm{tol}}$ is the tolerance for stress. The Newton correction in Equation (\ref{eqn:newton_correction}) is accepted if, 
\begin{equation}
\label{eqn:slip_converge}
\max_{\,\alpha} \; \lvert \Delta {\gamma_\textrm{p}}^{\alpha}\big(\mathbf{T}_{n+1}^{\textrm{e}}(\tau), \, s^{\alpha}_k(\tau) \big) \rvert < \Delta {\gamma_\textrm{p,\,tol}}, 
\end{equation}
where $\Delta {\gamma_\textrm{p,\,tol}}$ is the upper bound for the incremental shear strain rate. Here, we have used $\Delta {\gamma_\textrm{p,\,tol}}$ = 0.5. However, if the constraint in Equation (\ref{eqn:slip_converge}) is not satisfied, the elastic stress ($\mathbf{T}_{n+1}^{\textrm{e}}$) is further corrected by,
\begin{equation}
[T_{n+1}^{\textrm{e}}(\tau)]_{ij} = [T_{n}^{\textrm{e}}(\tau)]_{ij} + \eta \Delta T^{\textrm{e}}_{ij},
\end{equation}
where $\Delta \mathbf{T}^{\textrm{e}} = - \mathrsfso{J}^{-1}_n [\mathbf{G}_{\textrm{n}}]$ and $\eta$ is the correction factor. Here, we have used $\eta$ = 0.25. This correction is repeated until the corrected elastic stress satisfies the constraint.

Using the converged $\mathbf{T}^{\textrm{e}} (\tau)$, the slip resistance ($s^{\alpha}(\tau)$) and the dislocation density ($\rho^{\alpha}(\tau)$) are simply updated with no iterations by,
\begin{equation}
\begin{aligned}
s^{\alpha}_{k+1}(\tau) &= s^{\alpha}(t) + \Delta s^{\alpha}\Big( \Delta \gamma_\textrm{p}^{\beta} \big(\mathbf{T}_{n+1}^{\textrm{e}}(\tau), \, s^{\beta}_k(\tau) \big),\, \rho^{\beta}_k  \Big) \\
&= s^{\alpha}(t) + \frac{1}{2} \mu \frac{\sum_{\beta} a^{\alpha \beta}}{\sqrt{\sum_{\beta} \; a^{\alpha \beta} \rho^{\beta}_k}} \Bigg(\sqrt{\sum_{\gamma} d^{\beta \gamma} \rho^{\gamma}_k} - 2 y_c^{\beta} \rho^{\beta}_k \Bigg) \big\lvert \Delta \gamma_\textrm{p}^{\beta} \big(\mathbf{T}_{n+1}^{\textrm{e}}(\tau), \, s^{\beta}_k(\tau) \big) \big\rvert,
\end{aligned}
\end{equation}
and
\begin{equation}
\begin{aligned}
\rho^{\alpha}_{k+1}(\tau) &= \rho^{\alpha}(t) + \Delta \rho^{\alpha}\Big( \Delta \gamma_\textrm{p}^{\alpha} \big(\mathbf{T}_{n+1}^{\textrm{e}}(\tau), \, s^{\alpha}_k(\tau) \big),\, \rho^{\beta}_k  \Big)
\\&= \rho^{\alpha}(t) + \frac{1}{b} \Bigg( \sqrt{\sum_{\beta} d^{\alpha \beta} \rho^{\beta}_k} - 2 y_c^{\alpha} \rho^{\alpha}_k \Bigg) \big\lvert \Delta \gamma_\textrm{p}^{\alpha}\big(\mathbf{T}_{n+1}^{\textrm{e}}(\tau), \, s^{\beta}_k(\tau) \big) \big\rvert . 
\end{aligned}
\end{equation}
The slip resistance is accepted if,
\begin{equation}
\max_{\alpha} \lvert s^{\alpha}_{k+1} - s^{\alpha}_{k} \rvert < s^{\alpha}_{\textrm{tol}}.
\end{equation}
If not accepted, we go back to the first level of the iteration procedure in Equation (\ref{eqn:newton_correction}) upon using the updated values, $s^{\alpha}_{k+1}$ and $\rho^{\alpha}_{k+1}$.

Once $\mathbf{T}^\textrm{e} (\tau)$, $s^{\alpha}(\tau)$, $\rho^{\alpha}(\tau)$ in the two-step iteration procedure are accepted, the kinematic variables are then updated. The Cauchy stress and the temperature at the end of the increment are then updated by Equations (\ref{eqn:cauchy-2pk}) and (\ref{eqn:temperature}), respectively.

\section{Computation of material Jacobian}
\label{appendix:Jac}
\setcounter{equation}{0}

The implicit finite element procedure employed in this work uses a Newton-type iteration which requires a fourth order tangent also known as Jacobian at the end of the increment defined by,
\begin{equation}
\label{Definition_of_Jacobian}
\mathbfcal{W}(\tau) \equiv \frac{\partial \mathbf{T} (\tau)}{\partial \mathbf{E}_t (\tau)},
\end{equation}
where $\mathbf{T} (\tau)$ is the Cauchy stress and $\mathbf{E}_t (\tau)$ is the symmetric relative strain tensor\footnote[4]{The relative deformation gradient is defined as $\mathbf{F}_t(\tau) = \mathbf{F}(\tau) \mathbf{F}^{-1}(t)$. Similar to the deformation gradient, the relative deformation gradient tensor also allows for the polar decompostion, $\mathbf{F}_t(\tau) = \mathbf{R}_t(\tau) \mathbf{U}_t(\tau)$ where $\mathbf{R}_t$ is relative rotation tensor and $\mathbf{U}_t$ is relative stretch tensor. The relative strain tensor is then defined as $\mathbf{E}_t(\tau) = \textrm{ln} \mathbf{U}_t(\tau)$. (\cite{balasubramanian1998})}. The Cauchy stress is calculated by,
\begin{equation}
\label{eqn:Jac_true_elastic}
\mathbf{T} (\tau) = \frac{1}{\mathrm{det} \mathbf{F}^{\textrm{e}}(\tau)} [\mathbf{F}^{\textrm{e}}(\tau) \mathbf{T}^{\textrm{e}}(\tau) \mathbf{F}^{\textrm{e} \textrm{T}}(\tau) ].
\end{equation}
From Equation (\ref{eqn:Jac_true_elastic}) we have,
\begin{equation}
d\mathbf{T} = \frac{1}{\mathrm{det} \mathbf{F}^{\textrm{e}}} [d\mathbf{F}^{\textrm{e}} \mathbf{T}^{\textrm{e}} \mathbf{F}^{\textrm{e} \textrm{T}} +\mathbf{F}^{\textrm{e}} d\mathbf{T}^{\textrm{e}} \mathbf{F}^{\textrm{e} \textrm{T}}+ \mathbf{F}^{\textrm{e}} \mathbf{T}^{\textrm{e}} d\mathbf{F}^{\textrm{e} \textrm{T}}-(\mathbf{F}^{\textrm{e}} \mathbf{T}^{\textrm{e}} \mathbf{F}^{\textrm{e} \textrm{T}})\textrm{tr}(d\mathbf{F}^{\textrm{e}} \mathbf{F}^{\textrm{e} -1})].
\end{equation}
Hence, the tangent tensor is expressed by,
\begin{equation}
\mathcal{W}_{ijkl}= \frac{1}{\mathrm{det} \mathbf{F}^{\textrm{e}}} [\mathcal{S}_{imkl} T_{mn}^{\textrm{e}} F_{nj}^{\textrm{e} \textrm{T}} + F_{im}^{\textrm{e}} \mathcal{Q}_{mnkl} F_{nj}^{\textrm{e} \textrm{T}}+ F_{im}^{\textrm{e}} T_{mn}^{\textrm{e}} \mathcal{S}_{jnkl}-F_{im}^{\textrm{e}} T_{mn}^{\textrm{e}} F_{nj}^{\textrm{e} \textrm{T}}\mathcal{S}_{pqkl} F_{qp}^{\textrm{e} -1}],
\end{equation}
with 
\begin{equation}
\mathbfcal{S} \equiv \frac{\partial \mathbf{F}^{\textrm{e}}}{\partial \mathbf{E}_t} \quad \textrm{and} \quad \mathbfcal{Q} \equiv \frac{\partial \mathbf{T}^{\textrm{e}}}{\partial \mathbf{E}_t}.
\end{equation}
Since the relative stretch is small in this work,
\begin{equation}
\mathbf{E}_t = \textrm{ln} \mathbf{U}_t \approx \mathbf{U}_t -\mathbf{1},
\end{equation}
Where $\mathbf{U}_t$ is the relative stretch tensor. Therefore, $d\mathbf{E}_t \approx d\mathbf{U}_t$ and the fourth order tensor $\mathbfcal{S}$ and $\mathbfcal{Q}$ is expressed by,
\begin{equation}
\mathbfcal{S} = \frac{\partial \mathbf{F}^{\textrm{e}}}{\partial \mathbf{U}_t} \quad \textrm{and} \quad \mathbfcal{Q} = \frac{\partial \mathbf{T}^{\textrm{e}}}{\partial \mathbf{U}_t}.
\end{equation}

\noindent
(1) Calculation of $\mathbfcal{S}$\\

The elastic deformation gradient at $\tau$ can be obtained by,
\begin{equation}
\mathbf{F}^{\textrm{e}}(\tau) = \mathbf{F}(\tau) \mathbf{F}^{\textrm{p} -1}(\tau) = \mathbf{R}_t \mathbf{U}_t \mathbf{F}^{\textrm{e}}(t) \Bigg\{  \mathbf{1} - \sum_{\alpha=1}^{N} \Delta {\gamma_\textrm{p}}^{\alpha} \mathbf{\mathbb{S}}^{\alpha}_0 \Bigg\}.
\end{equation}
Then,
\begin{equation}
\begin{aligned}
\mathcal{S}_{ijkl} = \frac{\partial F_{ij}^{\textrm{e}}}{\partial U_{(t)\, kl}} &= \frac{\partial}{\partial U_{(t)\, kl}} \Bigg[R_{(t)\, im} U_{(t) \,mn} F_{np}^{\textrm{e}}(t) \Bigg\{  \delta_{pj} - \sum_{\alpha=1}^{N} \Delta {\gamma_\textrm{p}}^{\alpha} \mathbf{\mathbb{S}}^{\alpha}_{0 \, pj} \Bigg\}  \Bigg]\\
&= R_{(t)\, ik} F_{lj}^{\textrm{e}}(t) - R_{(t)\, ik} F_{lp}^{\textrm{e}}(t)\sum_{\alpha=1}^{N} \Delta {\gamma_\textrm{p}}^{\alpha} \mathbf{\mathbb{S}}^{\alpha}_{0 \, pj} - R_{(t)\, im} U_{(t) \,mn} F_{np}^{\textrm{e}}(t) \sum_{\alpha=1}^{N} \mathcal{R}_{kl}^{\alpha} \mathbf{\mathbb{S}}^{\alpha}_{0 \, pj},
\end{aligned}
\end{equation}
with $\mathbfcal{R}^{\alpha} = \frac{\partial \Delta {\gamma_\textrm{p}}^{\alpha}}{\partial \mathbf{U}_t}$, and subscript $(t)$ denotes relative quantities.
\\

\noindent
(2) Calculation of $\mathbfcal{Q}$\\

From Equation (\ref{eqn:stress_vs_trial_stress})
\begin{equation}
\label{eqn:jacob_Q_tensor}
\mathcal{Q}_{ijkl} = \frac{\partial T_{ij}^{\textrm{e}}}{\partial U_{(t) \, kl}} = \mathcal{D}_{ijkl} - \sum_{\alpha=1}^{N} \mathcal{R}_{kl}^{\alpha} C_{ij}^{\alpha} - \sum_{\alpha=1}^{N} \Delta {\gamma_\textrm{p}}^{\alpha} \mathcal{J}^{\alpha}_{ijkl},
\end{equation}
where $\mathbfcal{D} = \frac{\partial \mathbf{T}^{\textrm{e}}_{\textrm{tr}}}{\partial \mathbf{U}_t}$ and $\mathbfcal{J}^{\alpha}=\frac{\partial \mathbf{C}^{\alpha}}{\partial \mathbf{U}_t}$. $\mathbfcal{D}$ is expressed by,
\begin{equation}
\mathcal{D}_{ijkl} = \frac{\partial T_{\textrm{(tr)} \, ij}^{\textrm{e}}}{\partial U_{(t) \, kl}} = \frac{1}{2} \mathcal{C}_{ijmn} \mathcal{L}_{mnkl}.
\end{equation}
And $\mathbfcal{L}$ is calculated by,
\begin{equation}
\begin{aligned}
\mathcal{L}_{ijkl} = \frac{\partial C_{\textrm{(tr)} \, ij}^{\textrm{e}}}{\partial U_{(t) \, kl}} &= \frac{\partial}{\partial U_{(t) \, kl}}\Big[ F_{mi}^{p -1}(t) F_{nm}(\tau) F_{np}(\tau) F_{pj}^{p -1}(t) \Big]\\
&=\frac{\partial}{\partial U_{(t) \, kl}}\Big[ F_{mi}^e(t) U_{(t) \, mn}(\tau) U_{(t) \, np}(\tau) F_{pj}^e(t)\Big]\\
&= F_{mi}^e(t) \delta_{mk} \delta_{nl} U_{(t) \, np} F_{pj}^e(t) + F_{mi}^e(t) U_{(t) \, mn} \delta_{nk} \delta_{pl} F_{pj}^e(t)\\
&= F_{ki}^e(t) U_{(t) \, lp} F_{pj}^e(t) + F_{mi}^e(t) U_{(t) \, mk} F_{lj}^e(t).
\end{aligned}
\end{equation}
Furthermore, from Equations (\ref{eqn:b_alpah}) and (\ref{eqn:c_alpah}), $\mathbfcal{J}^{\alpha}$ is calculated by,
\begin{equation}
\begin{aligned}
\mathcal{J}^{\alpha}_{ijkl}=\frac{\partial C_{ij}^{\alpha}}{\partial U_{(t) \, kl}} &= \frac{\partial}{\partial U_{(t) \, kl}} \Bigg[\frac{1}{2} \mathcal{C}_{ijmn} \big( C_{\textrm{(tr)} \, mp} \mathbb{S}^{\alpha}_{0 \, pn } + \mathbb{S}^{\alpha}_{0 \, pm }C_{\textrm{(tr)} \, pn} \big) \Bigg]\\
&= \frac{1}{2} \Big[ \mathcal{C}_{ijmn} \mathcal{L}_{mpkl} \mathbb{S}^{\alpha}_{0 \, pn } + \mathcal{C}_{ijmn}\mathbb{S}^{\alpha}_{0 \, pm } \mathcal{L}_{pnkl} \Big].
\end{aligned}
\end{equation}
Moreover, $\mathbfcal{R}^{\alpha}$ is calculated by,
\begin{equation}
\label{eqn:jacob_R_tensor}
\mathcal{R}_{ij}^{\alpha} = \frac{\partial \Delta \gamma_\textrm{p}^{\alpha} (\mathbf{T}^e)}{\partial U_{(t) \, ij}} = \frac{\partial \Delta \gamma_\textrm{p}^{\alpha} (\mathbf{T}^e)}{\partial T_{kl}^{\textrm{e}}} \frac{{\partial T_{kl}^{\textrm{e}}}}{\partial U_{(t) \, ij}} = \mathcal{B}^{\alpha}_{kl} \mathcal{Q}_{klij},
\end{equation}
with
\begin{equation}
\mathcal{B}^{\alpha}_{ij} = \frac{\partial \Delta \gamma_\textrm{p}^{\alpha}}{\partial {T}_{ij}^e} = \frac{\partial \Delta \gamma_\textrm{p}^{\alpha}}{\partial \tau^{\alpha}} \frac{\partial \tau^{\alpha}}{\partial {T}_{ij}^e} = \frac{\partial \Delta \gamma_\textrm{p}^{\alpha}}{\partial \tau^{\alpha}} \frac{1}{2} (\mathbb{S}^{\alpha}_{0 \, ij} + \mathbb{S}^{\alpha}_{0 \, ji}).
\end{equation}
$\mathbfcal{Q}$ is  therefore expressed by,
\begin{equation}
\label{eqn:jacab_q_rearrange}
\mathbfcal{Q} = \mathbfcal{D} - \sum_{\alpha=1}^{N} (\mathbf{C}^{\alpha} \otimes \mathbfcal{B}^{\alpha}) \mathbfcal{Q} -\sum_{\alpha=1}^{N} \Delta {\gamma_\textrm{p}}^{\alpha} \mathbfcal{J}^{\alpha}.
\end{equation}
Then, rearranging Equation (\ref{eqn:jacab_q_rearrange}), we have,
\begin{equation}
\label{jacob_Q_final}
\mathbfcal{Q} = \Bigg[ \mathbfcal{I} + \sum_{\alpha=1}^{N} (\mathbf{C}^{\alpha} \otimes \mathbfcal{B}^{\alpha}) \Bigg]^{-1} \Bigg[ \mathbfcal{D} -\sum_{\alpha=1}^{N} \Delta {\gamma_\textrm{p}}^{\alpha} \mathbfcal{J}^{\alpha} \Bigg],
\end{equation}
\begin{equation}
\label{jacob_K}
\mathbfcal{K} \equiv \mathbfcal{I} + \sum_{\alpha=1}^{N} (\mathbf{C}^{\alpha} \otimes \mathbfcal{B}^{\alpha}),
\end{equation}
\begin{equation}
\label{jacob_M}
\mathbfcal{M} \equiv \mathbfcal{D} -\sum_{\alpha=1}^{N} \Delta {\gamma_\textrm{p}}^{\alpha} \mathbfcal{J}^{\alpha}.
\end{equation}

The computation of the tangent is summarized as follows,\\
\begin{enumerate}
\item $\mathbf{U}_{t} =\mathbf{R}_{t}^{-1} \mathbf{F}(\tau) \mathbf{F}^{-1}(t)   $,
\item $\mathcal{C}_{ijkl} = Q_{ip} Q_{jq} Q_{kr} Q_{lr} \mathcal{C}^{c}_{pqrs},$
\item $\mathcal{L}_{ijkl} = F_{ki}^e(t) U_{(t) \, lp} F_{pj}^e(t) + F_{mi}^e(t) U_{(t) \, mk} F_{lj}^e(t)$,
\item $\mathcal{D}_{ijkl} = \frac{1}{2} \mathcal{C}_{ijmn} \mathcal{L}_{mnkl}$,
\item $\mathcal{J}^{\alpha}_{ijkl} = \frac{1}{2} \Big[ \mathcal{C}_{ijmn} \mathcal{L}_{mpkl} \mathbb{S}^{\alpha}_{0 \, pn } + \mathcal{C}_{ijmn}\mathbb{S}^{\alpha}_{0 \, pm } \mathcal{L}_{pnkl} \Big]$,
\item $\mathcal{B}^{\alpha}_{ij} = \frac{\partial \Delta \gamma_\textrm{p}^{\alpha}}{\partial \tau^{\alpha}} \frac{1}{2} (\mathbb{S}^{\alpha}_{0 \, ij} + \mathbb{S}^{\alpha}_{0 \, ji})$,
\item $\mathcal{K}_{ijkl} = \delta_{ik}\delta_{jl} + \sum_{\alpha=1}^{N} C_{ij}^{\alpha} \mathcal{B}_{kl}^{\alpha}$, (Calculated in reduced form $\mathcal{K}_{IJ}$)
\item $\mathcal{M}_{ijkl} = \mathcal{D}_{ijkl} -\sum_{\alpha=1}^{N} \Delta {\gamma_\textrm{p}}^{\alpha} \mathcal{J}_{ijkl}^{\alpha}$. (Calculated in reduced form $\mathcal{M}_{IJ}$)
\item $\mathcal{Q}_{ijkl} = \mathcal{K}^{-1}_{ijmn} \mathcal{M}_{mnkl}$ (Calculated in reduced form $\mathcal{Q}_{IJ}$)
\item $\mathcal{R}_{ij}^{\alpha} = \mathcal{B}^{\alpha}_{kl} \mathcal{Q}_{klij}$,
\item $\mathcal{S}_{ijkl} = R_{(t)\, ik} F_{lj}^{\textrm{e}} - R_{(t)\, ik} F_{lp}^{\textrm{e}}\sum_{\alpha=1}^{N} \Delta {\gamma_\textrm{p}}^{\alpha} \mathbf{\mathbb{S}}^{\alpha}_{0 \, pj} - R_{(t)\, im} U_{(t) \,mn} F_{np}^{\textrm{e}} \sum_{\alpha=1}^{N} \mathcal{R}_{kl}^{\alpha} \mathbf{\mathbb{S}}^{\alpha}_{0 \, pj},$
\item $\mathcal{W}_{ijkl}= \frac{1}{\mathrm{det} \mathbf{F}^{\textrm{e}}} [\mathcal{S}_{imkl} T_{mn}^{\textrm{e}} F_{nj}^{\textrm{e} \textrm{T}} + F_{im}^{\textrm{e}} \mathcal{Q}_{mnkl} F_{nj}^{\textrm{e} \textrm{T}}+ F_{im}^{\textrm{e}} T_{mn}^{\textrm{e}} \mathcal{S}_{jnkl}-F_{im}^{\textrm{e}} T_{mn}^{\textrm{e}} F_{nj}^{\textrm{e} \textrm{T}}\mathcal{S}_{pqkl} F_{qp}^{\textrm{e} -1}]$
\end{enumerate}

\end{appendices}
\clearpage
\printbibliography

\end{document}